\documentclass[superscriptaddress,prd,twocolumn,showpacs,preprintnumbers,amsmath,amssymb,nobibnotes,secnumarabic,altaffilletter,floatfix,aps,nofootinbib,10pt]{revtex4-1}
\usepackage{graphicx}
\usepackage{amsmath}
\usepackage{amsfonts}
\usepackage{float}
\usepackage{bbold}
\usepackage{textcomp}
\usepackage{xcolor}
\usepackage{ulem}
\usepackage[T1]{fontenc}
\usepackage[utf8]{inputenc}
\usepackage{soul}

\usepackage{lineno}
\usepackage{multirow}

\begin{document}

\pagestyle{plain}


\title{A search for cosmogenic production of $\beta$-neutron emitting radionuclides in water}
\author{S. Dazeley}
\email[email: ]{dazeley2@llnl.gov}
\affiliation{Lawrence Livermore National Laboratory, Livermore, CA 94550, USA}
\author{M. Askins}
\affiliation{Physics Department, University of California, Davis, CA 95616, USA}
\author{M. Bergevin}
\affiliation{Lawrence Livermore National Laboratory, Livermore, CA 94550, USA}
\author{A. Bernstein}
\affiliation{Lawrence Livermore National Laboratory, Livermore, CA 94550, USA}
\author{N. S. Bowden}
\affiliation{Lawrence Livermore National Laboratory, Livermore, CA 94550, USA}
\author{P. Jaffke}
\affiliation{Center for Neutrino Physics, Virginia Tech, Blacksburg, VA 24061, USA}
\author{S.D. Rountree}
\affiliation{Center for Neutrino Physics, Virginia Tech, Blacksburg, VA 24061, USA}
\author{T.M. Shokair}
\affiliation{Department of Nuclear Engineering, University of California, Berkeley, CA 94720, USA}
\author{M. Sweany}
\altaffiliation[Currently at: ]{Sandia National Laboratories, Livermore, CA}
\affiliation{Lawrence Livermore National Laboratory, Livermore, CA 94550, USA}

\begin{abstract}
Here we present the first results of WATCHBOY, a water Cherenkov detector designed to measure the yield of $\beta$-neutron emitting radionuclides produced by cosmic ray muons in water. In addition to the $\beta$-neutron measurement, we also provide a first look at isolating single-$\beta$ producing radionuclides following muon-induced hadronic showers as a check of the detection capabilities of WATCHBOY. The data taken over $207$ live days indicates a $^{9}$Li production yield upper limit of $1.9\times10^{-7}\mu^{-1}g^{-1}\mathrm{cm}^2$ at $\sim400$ meters water equivalent (m.w.e.) overburden at the $90\%$ confidence level. In this work the $^{9}$Li signal in WATCHBOY was used as a proxy for the combined search for $^{9}$Li and $^{8}$He production. This result will provide a constraint on estimates of antineutrino-like backgrounds in future water-based antineutrino detectors.
\end{abstract}

\maketitle

%
%

\section{Introduction}

In recent years a number of water-based detector concepts sensitive to reactor antineutrinos through the inverse $\beta$ decay (IBD) reaction have been proposed~\cite{beacom2004antineutrino,Adams:2013qkq,Askins:2015bmb,Gann:2015fba}. Unlike organic scintillator, water has better light propagation properties, is more benign environmentally, and can be more cost effective as detectors get larger. Furthermore, the advent of water-based liquid scintillator (WBLS)~\cite{yeh2011new} offers the possibility of hybrid scintillator/Cherenkov detectors capable of directional charged particle sensitivity, efficient neutron tagging, low-energy-thresholds and excellent energy resolution for neutrino, double-$\beta$-decay, and proton-decay experiments~\cite{alonso2014advanced}. 

Cosmic-ray muon spallation products are potential sources of backgrounds in such detectors. Radionuclide production via muon initiated spallation in organic liquid scintillator has been studied extensively in various detectors worldwide at varying depths~\cite{Eguchi:2002dm,Abe:2011fz,Alimonti:2000xc,An:2012eh,Ahn:2012nd}. For antineutrino experiments, the most dangerous radionuclides are long-lived isotopes that decay via simultaneous emission of a $\beta$ and a neutron ($\beta$-neutron), such as $^{9}$Li and $^{8}$He. In principle these isotopes, and some others, can also be formed in water, and could contribute significantly to antineutrino backgrounds in water-based antineutrino detectors. Recently, the rates of $^{9}$Li and $^{8}$He production in water were measured for the first time at Super-Kamiokande (SK)~\cite{SKRadNucs2015}. Though neutron tagging at SK is inefficient due to the lack of a neutron capture agent such as gadolinium, the large volume and extended data-taking period enabled a measurement. The result ($0.51 \pm 0.07 \pm 0.09 \times 10^{-7}\mu^{-1}g^{-1}\mathrm{cm}^2$), appears to be almost a factor four lower than the FLUKA-based predictions of Li and Beacom~\cite{Li:2014}. In the present study, we demonstrate a different approach, in which the neutron-tagging efficiency, and thus the efficiency for the $\beta$-neutron radionuclides of interest, is increased compared to SK, through the addition of a gadolinium dopant~\cite{Sweany2011,Watanabe2009}. SK has recently announced that it will make use of this gadolinium-doping technique in a planned upgrade \cite{SymMag}. In this paper, we present the first results from a water detector using a gadolinium tag to measure $\beta$-neutron radionuclides at a depth of approximately $400$ meters water equivalent (m.w.e.) at the Kimballton Underground Research Facility (KURF) in the Kimballton mine in Virginia~\cite{KURFsite}. The measurement is significant since it demonstrates the technique for use in larger detectors like SK, and because it is at a different depth, permitting a constraint on the uncertain depth scaling factors that are used to extrapolate results between different overburdens. 

\section{WATCHBOY detector}

\begin{figure}[ht]
\includegraphics[width=0.48\textwidth, height=0.30\textheight]{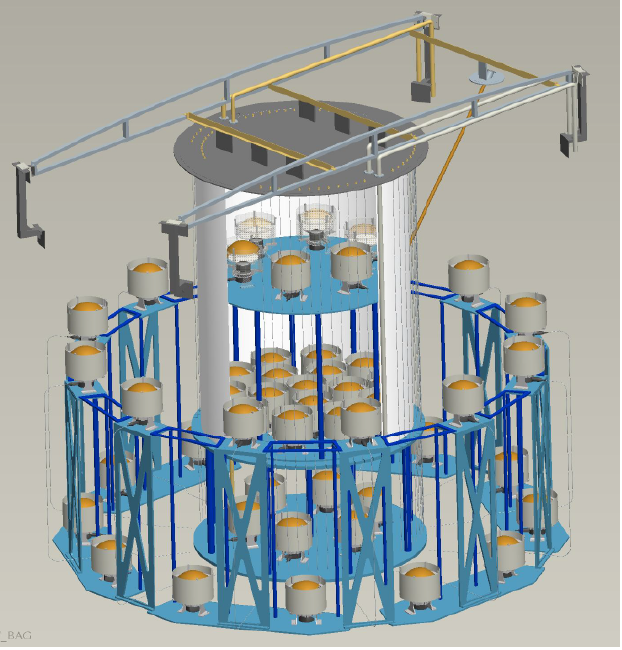}
\includegraphics[width=0.48\textwidth, height=0.30\textheight]{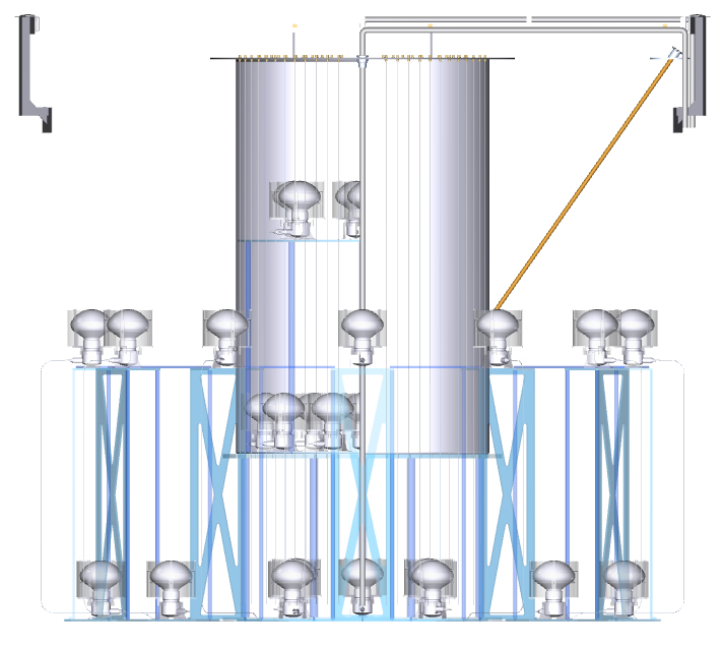}
\caption{\label{fig:watchboydesign} The PMT arrangement and supporting structure inside the WATCHBOY detector. The gadolinium doped target region containing $16$ tightly packed upward facing PMTs is shown at the center of the detector, visible through an illustrated cutout of the target containment bag, which physically and optically separates the inner target from the veto. The veto PMTs are mounted on the far outside and base. At the top, inside the bag is another optically separated region that forms part of the veto (this region of the veto contains Gd).}
\end{figure}

Conceived as a prototype to WATCHMAN~\cite{Askins:2015bmb}, a kiloton-scale reactor antineutrino detector proposed for the Morton salt mine, 13km from the Perry reactor, WATCHBOY was designed to measure the production yields of long-lived radionuclides that mimic the antineutrino induced IBD reaction in water-based media. WATCHBOY is a water-Cherenkov detector with a $\sim2$ ton target filled with pure deionized water plus $0.2\%$ GdCl$_3$. Natural gadolinium is an excellent neutron absorber, having a neutron capture cross-section of $49,000$ barns~\cite{PDG2012}. Upon capture, the nucleus emits a gamma ray shower summing to approximately $8\,\mathrm{MeV}$. At the base of the target, $16$ upward looking $10$" Hamamatsu R$7081$ photomutiplier tubes (PMTs) collect the Cherenkov light from particle interactions inside. The target walls are coated with a reflective Teflon based material (GORE$\,\circledR$ DRP$\,\circledR$) that maximizes light reflection and detection at the PMTs. Surrounding the target is a $\sim40$ ton pure water volume for identifying and tagging cosmic ray muons, instrumented by $36$ $10$" Hamamatsu R7081 PMTs. Fig.~\ref{fig:watchboydesign} shows a schematic of the WATCHBOY detector.

\begin{figure}[ht]
\centering 
\includegraphics[width=0.48\textwidth]{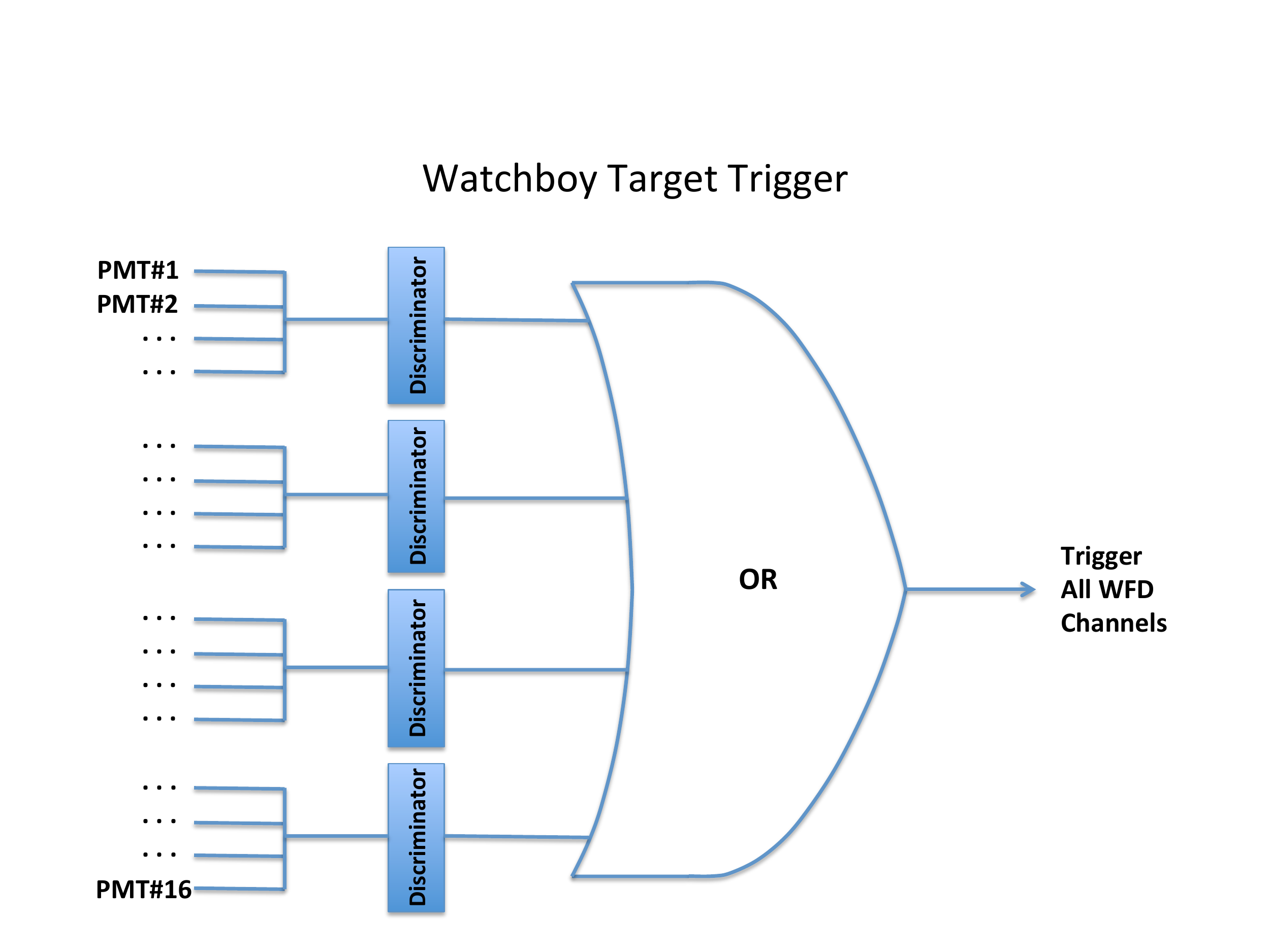}
\caption{The Watchboy target trigger logic.}
\label{fig:trigger}
\end{figure}

The signal processing and triggering scheme for WATCHBOY is done as follows. Each PMT signal is sent to one of four Struck SIS3316 sixteen-channel digitizer boards, with 250 MHz sampling rate and 14 bit dynamic range. The PMT signals on each board are grouped into sets of four, with the sum from each set sent to an onboard discriminator. There are four discriminators per board. If any discriminator is triggered, a signal is sent to a CAEN V1495 FPGA, which in turn sends a global trigger to all 52 channels to read-out to disk. Fig.~\ref{fig:trigger} shows a schematic of the trigger.

Since $^{9}$Li normally forms the major component of the $\beta$-neutron background in liquid scintillator based antineutrino experiments
\cite{DCBackgrounds2013,KamLANDbackgrounds}, and since both the $^{9}$Li and $^{8}$He $\beta$ energy spectra, and mean lifetimes are very similar ($^{8}$He $\tau = 172$ ms, Q value = 10.7 MeV) ($^{9}$Li $\tau$ = 257 ms, Q value = 11.9 MeV)~\cite{Tilley:2004zz}, $^{9}$Li will serve as a proxy for a combined search for both $^{9}$Li and $^{8}$He in WATCHBOY. To summarize the radionuclide signature in WATCHBOY, we search for a muon passing through the target, followed $\sim 257$ ms later by a correlated pair of events - a $\beta$ followed by a neutron capture.



\section{ Data Selection and Run Stability}
\begin{figure*}[ht]
\centering  
\includegraphics[width=0.49\textwidth, height=0.25\textheight]{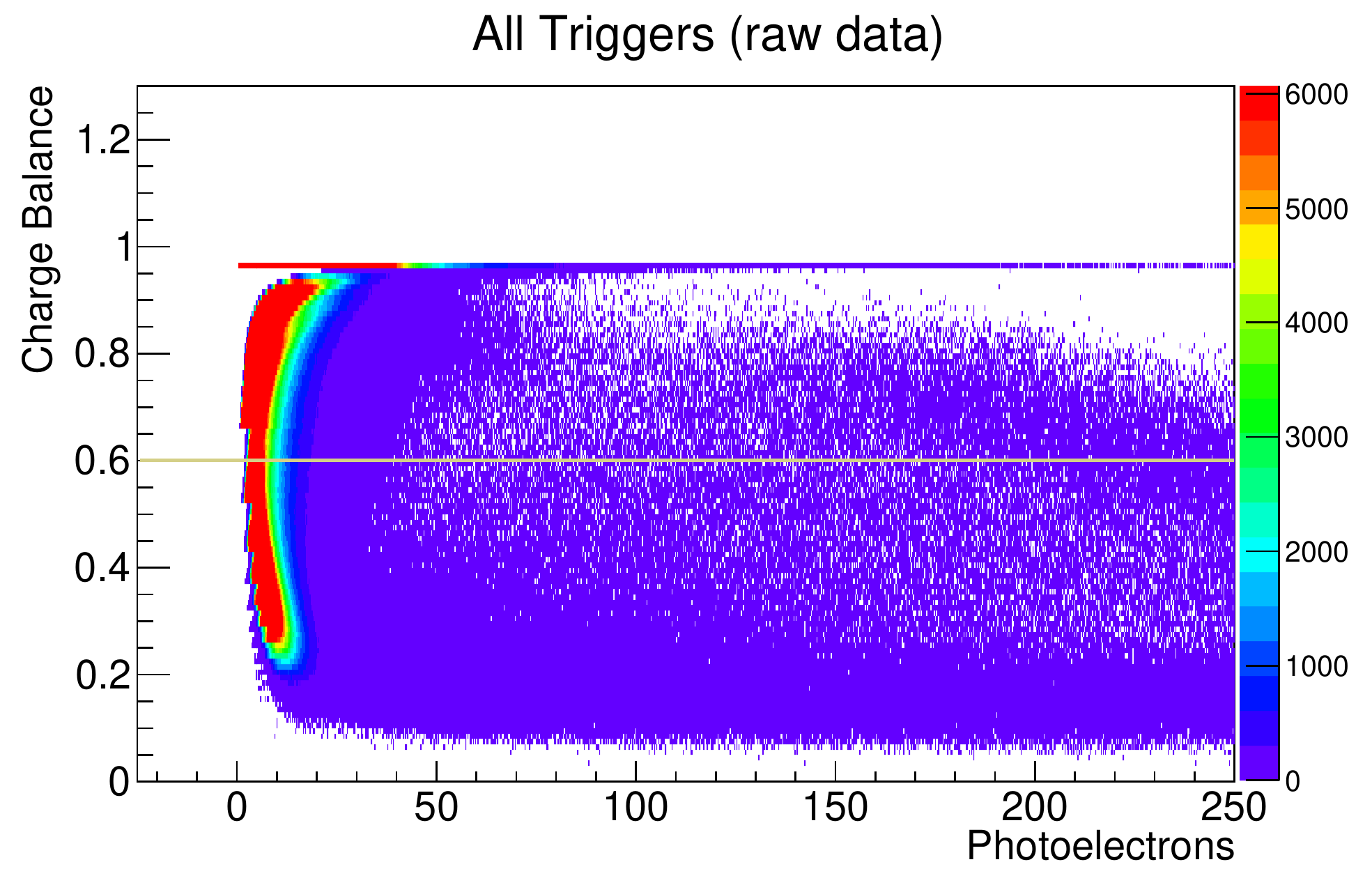}
\includegraphics[width=0.49\textwidth, height=0.25\textheight]{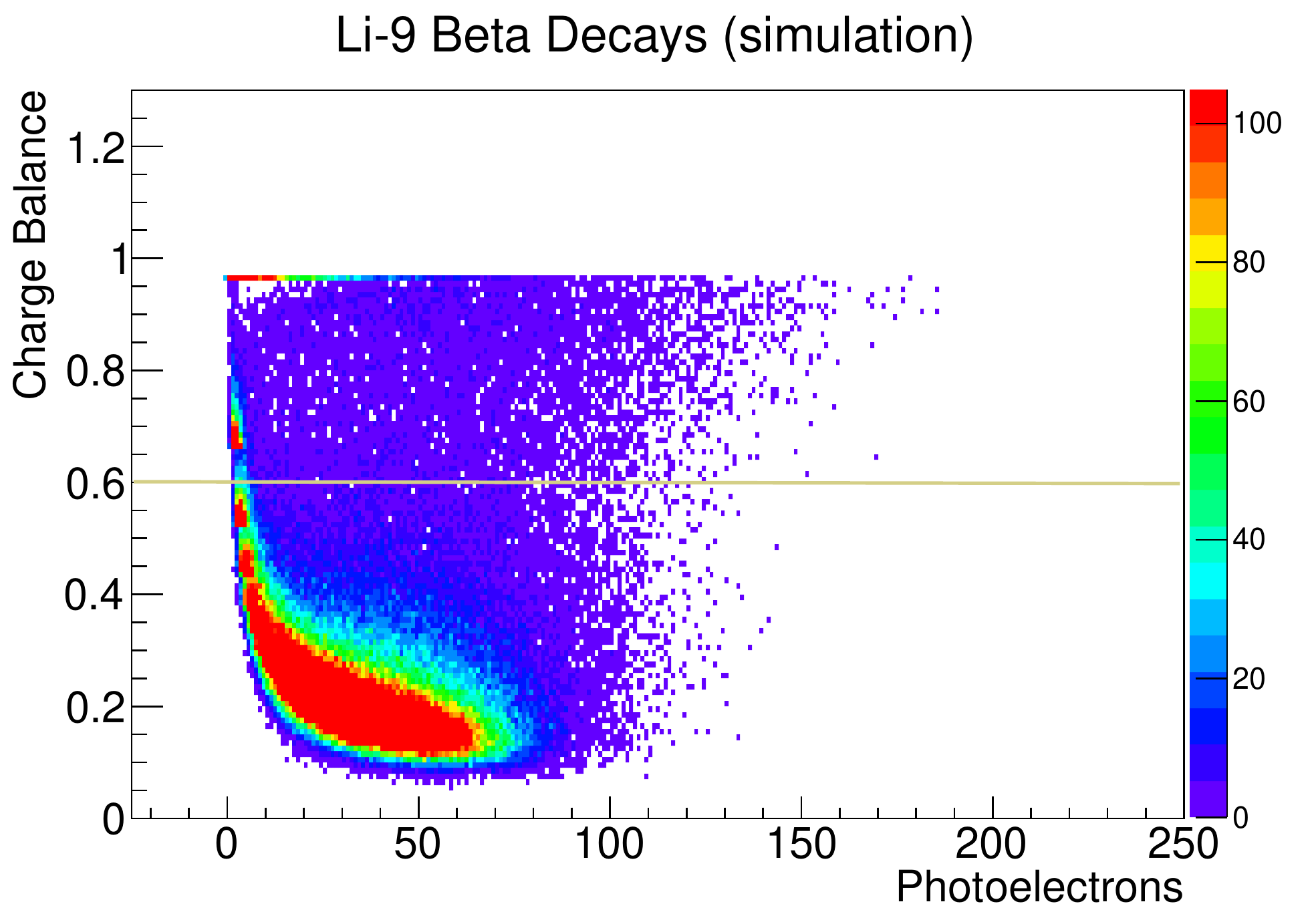}
\includegraphics[width=0.49\textwidth, height=0.25\textheight]{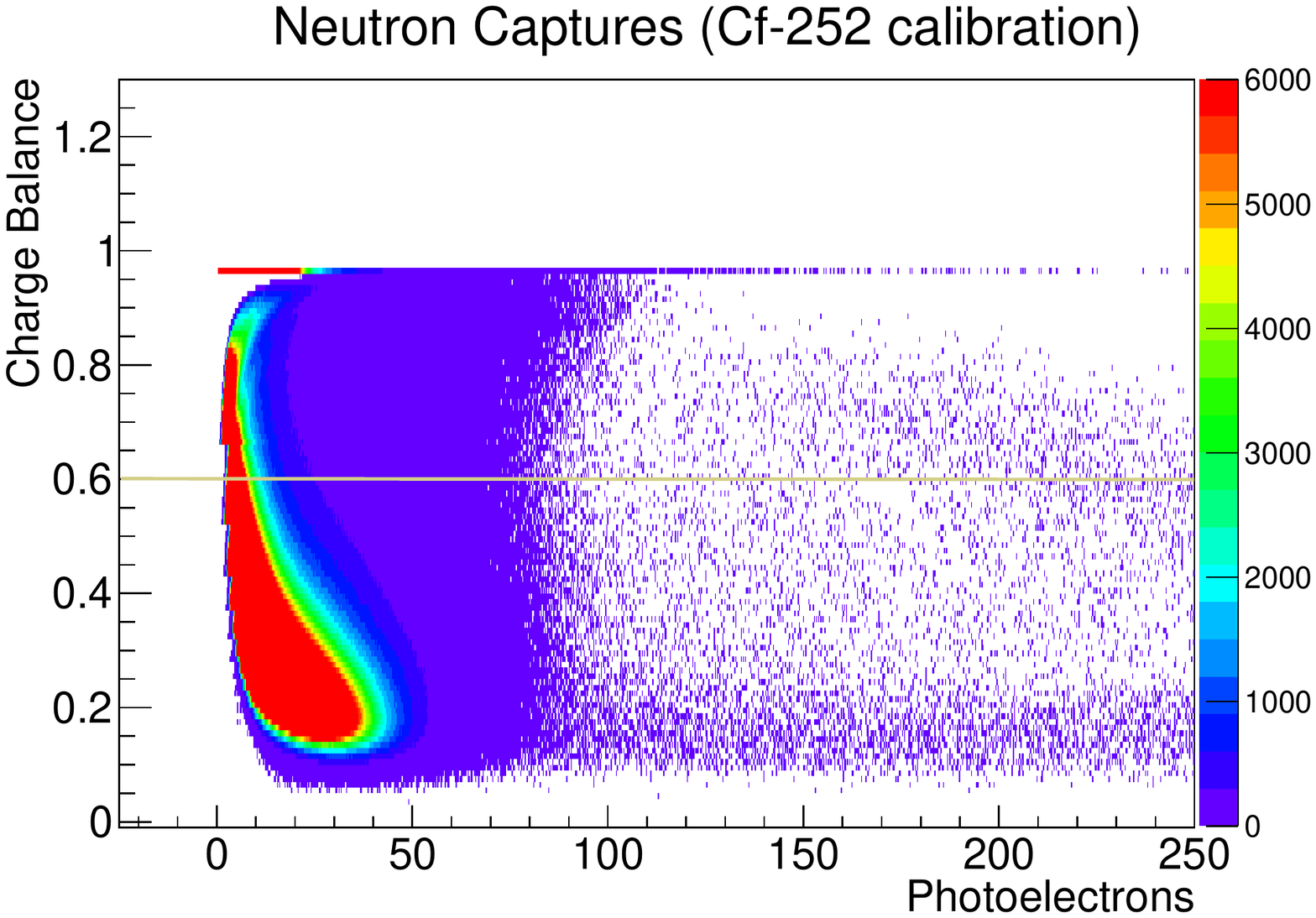}
\includegraphics[width=0.49\textwidth, height=0.25\textheight]{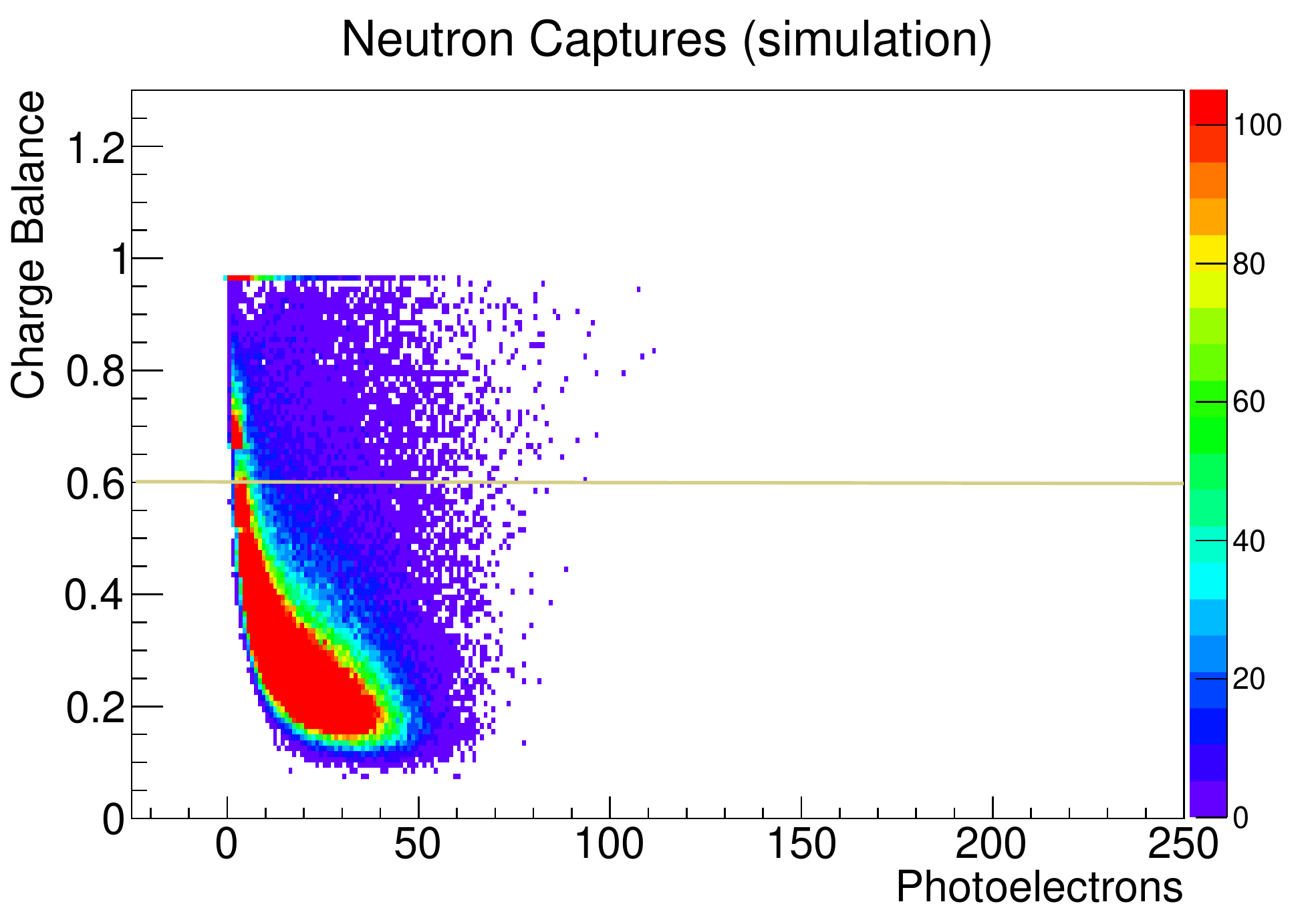}
\caption{\label{fig:chrgbalance} A comparison of the evenness of the light distribution of events in the WATCHBOY target for physics and neutron capture events (left). The distributions that result from the simulation of $^{9}$Li $\beta$-decay and neutron capture are shown for comparison (right). The evenness of the light distribution is measured using the ``Charge Balance'' parameter defined in Eq.~\ref{eq:chrgbalance}, low values result from an even light distribution among the PMTs, large values result from high concentrations of the signal in a small number of PMTs. The horizontal line at charge balance = 0.6 indicates the upper limit for $^{9}$Li candidate events in this analysis.}
\end{figure*}

The WATCHBOY detector began taking data in late July 2013. Early data was used to characterize and tune the data acquisition rates. The PMT gains were adjusted for the final time in September 2013 after an LED calibration.

A data selection criterion was implemented in order to reject the large rate of instrumental-noise events, while retaining the majority of the physics signal. Fig.~\ref{fig:chrgbalance} shows a scatter plot of total event charge versus a measure of the evenness of the light distribution among the target PMTs for physics and $^{252}$Cf calibration data. Simulation of neutron capture and $^{9}$Li beta decay in the target are also included for comparison. The evenness of the light distribution (charge balance) is defined as
\begin{equation}
\mathrm{Charge\,\,Balance}\,=  \sqrt{ \frac{\Sigma Q_i^{2}}{(Q_{sum})^2} - \frac{1}{N}}
\label{eq:chrgbalance}
\end{equation}
where $N$ is the number of PMTs, $Q_i$ is the charge of the $i^{th}$ PMT, and $Q_{\mathrm{sum}}$ is the summed charge of all the PMTs.
Events with an even distribution of light among all $16$ target PMTs will produce charge balance values approaching zero. Values close to one indicate the opposite extreme (i.e. most of the signal concentrated in one or two PMTs). Note that the neutron
capture and $^{9}$Li $\beta$ events tend towards low charge balance values, especially as the total charge increases. This effect is replicated by the simulation also. We therefore further require that genuine physics events of interest have a relatively even distribution parameter ($<0.6$). 

In addition to the charge balance cut, we define an ``Event of Interest'' as any event that passes the charge balance cut while producing a signal of between 13 and 100 photoelectrons. As we shall see in Section 4, the ``Event of Interest'' dataset is a superset containing all the genuine $^{9}$Li $\beta$-like or neutron-like candidates plus additional events that just failed to meet the strict criteria for a genuine candidate.
The rate of ``Events Of Interest'' from July 2013 until August 2014 varied between $1.6$ and $0.5$Hz.  Fig.~\ref{fig:eventRate} shows the rate of such events evolving over the data taking period. The drop in event rates around day $\#100$ was caused by the loss of a high voltage module and was not included in the analysis. The sudden drops in event rate around day $\#290$ and again at day $\#350$ were caused, in both cases, by the loss of target PMTs. Note that, after day $\#290$, the fractional loss of signal per neutron capture, as indicated by the end points of the neutron capture curve of Fig.~\ref{fig:eventRate}, have a disproportionally large impact on the overall trigger rate, since for the purposes of the trigger, the PMT signals are grouped into groups of four (the trigger scheme is shown in Fig.~\ref{fig:trigger}). The loss of a single PMT in a group reduces the trigger rate from the whole group. Losses in detector sensitivity might also have been expected due to gradual degradation of water quality over time. Some evidence of this was apparent between days $\#290$ and $\#350$ (a $\sim 8\%$ drop), and between $\#350$ and $\#410$ (a $\sim 6\%$ drop). However, the effect appears to be small relative to the abrupt PMT related losses already noted, indicating that water quality was relatively stable over the data period. 

We now describe the method used to track, and adjust for, such changes using the neutron capture candidates that sometimes follow muon spallation in the water. Since neutrons produced via muon spallation and spontaneous fission in a $^{252}$Cf source are correlated, either with the muon in the case of muon spallation, or with other neutrons in a $^{252}$Cf fission. the second event in a correlated pair is highly likely to be a neutron capture. An ensemble of such correlated pairs can be used to provide a statistically pure spectrum of neutron capture events. Note, in muon spallation, the first event is provided by the muon traversing the target. Pure neutron spectra can be obtained by searching for energy deposits correlated with the muon candidates. Fig.~\ref{fig:eventRate} shows the neutron capture spectrum accumulated from the period of data between day $\#140$ and $\#290$. The $^{252}$Cf calibrations shown earlier (Fig.~\ref{fig:2}), was also taken during this time. In red we show a Gaussian fit to the spectrum. Note that the shape of the neutron capture data is remarkably Gaussian. In order to quantify changes in the detector response over time we use this Gaussian fit to determine the number of photoelectrons corresponding to $3 \sigma$ above the Gaussian mean, which is used as a proxy for the detector response for a full 8 MeV energy deposit in the target. The black points {(with error bars) shown in Fig.~\ref{fig:eventRate} show the evolution of the $3 \sigma$ point over time. The number of photoelectrons corresponding to 8 MeV is used to adjust the energy cuts associated with $^{9}$Li $\beta$s and neutrons (and ``events of interest'') over the whole data taking period. Fig.~\ref{fig:eventRateAdjusted} shows the rate of ``Events of Interest'' once adjusted photoelectron cuts are applied. We note that the event rate stabilizes to within $\pm12\%$ after the application of the adjusted cuts. We also note that the adjusted data rates near the transitions (day numbers 100, 290 and 350) remain somewhat unstable, even with the correction applied, and are not included in the following analysis.
\begin{figure}[ht]
\centering
\includegraphics[width=0.49\textwidth, height=0.25\textheight]{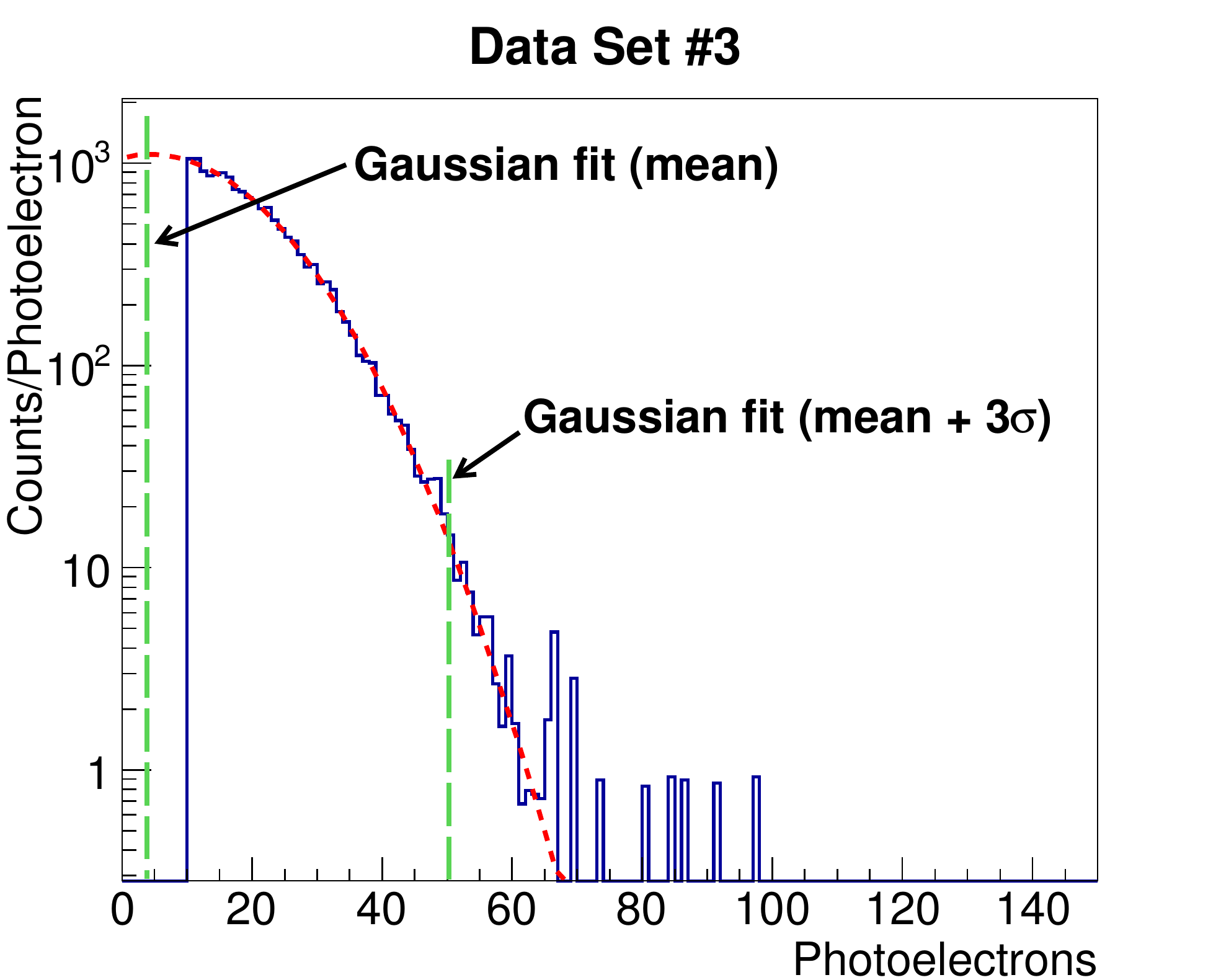}
\includegraphics[width=0.49\textwidth, height=0.25\textheight]{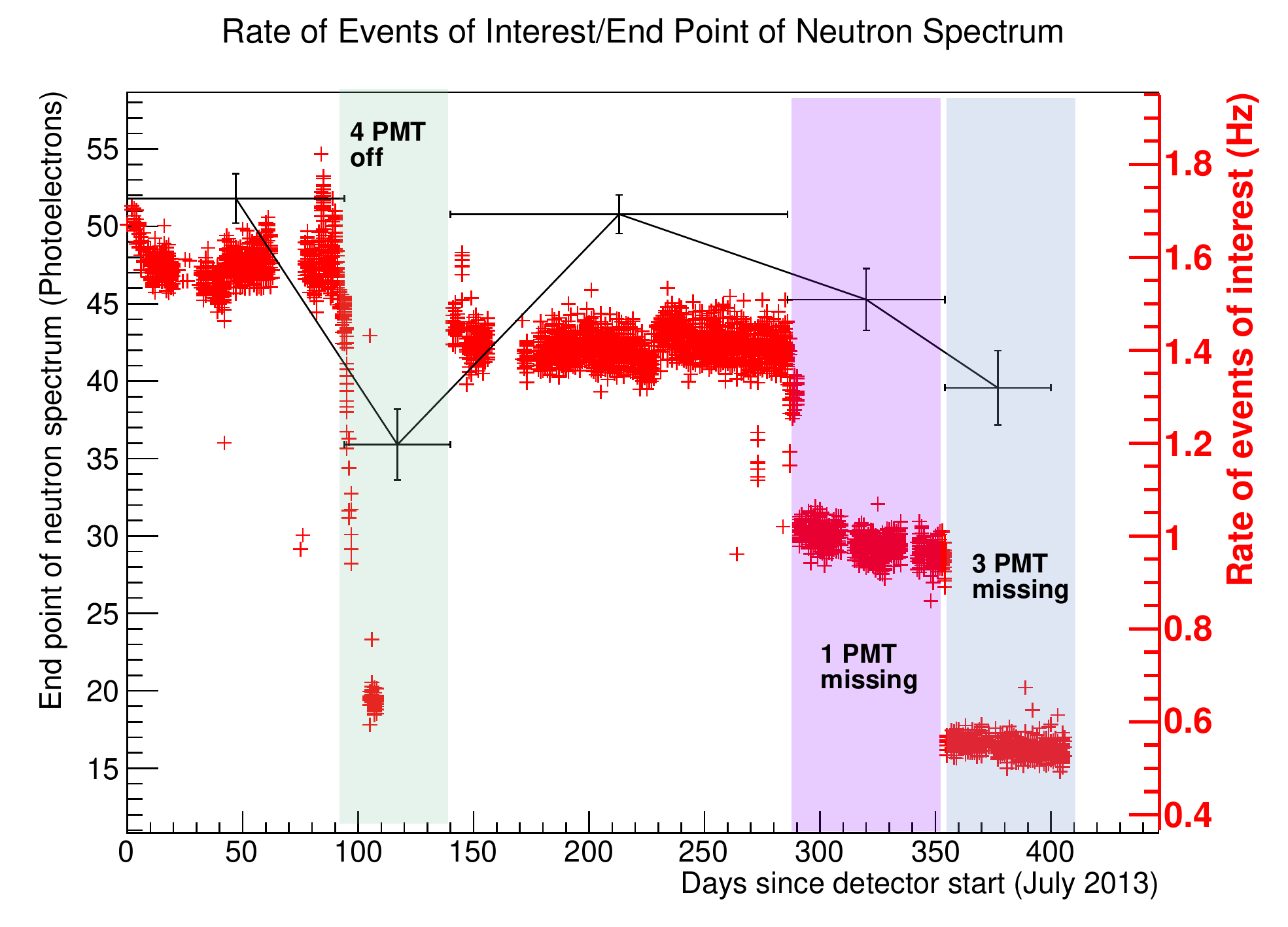}
\caption{(Top: the spectrum of all neutron capture events that result from muon spallation in the target from day $\#140$ to day $\#290$, together with the Gaussian fit. Also shown are the mean and the upper $3\sigma$ of the Gaussian fit. Bottom: the evolution of the rate of ``Events of Interest'' (described in text) over the course of the data taking period (red), together with the evolution of the number of photoelectrons corresponding to the Gaussian $3\sigma$ point, which approximates an 8 MeV energy deposition in the target (black).}
\label{fig:eventRate}
\end{figure}
\begin{figure}[ht]
\centering
\includegraphics[width=0.49\textwidth, height=0.25\textheight]{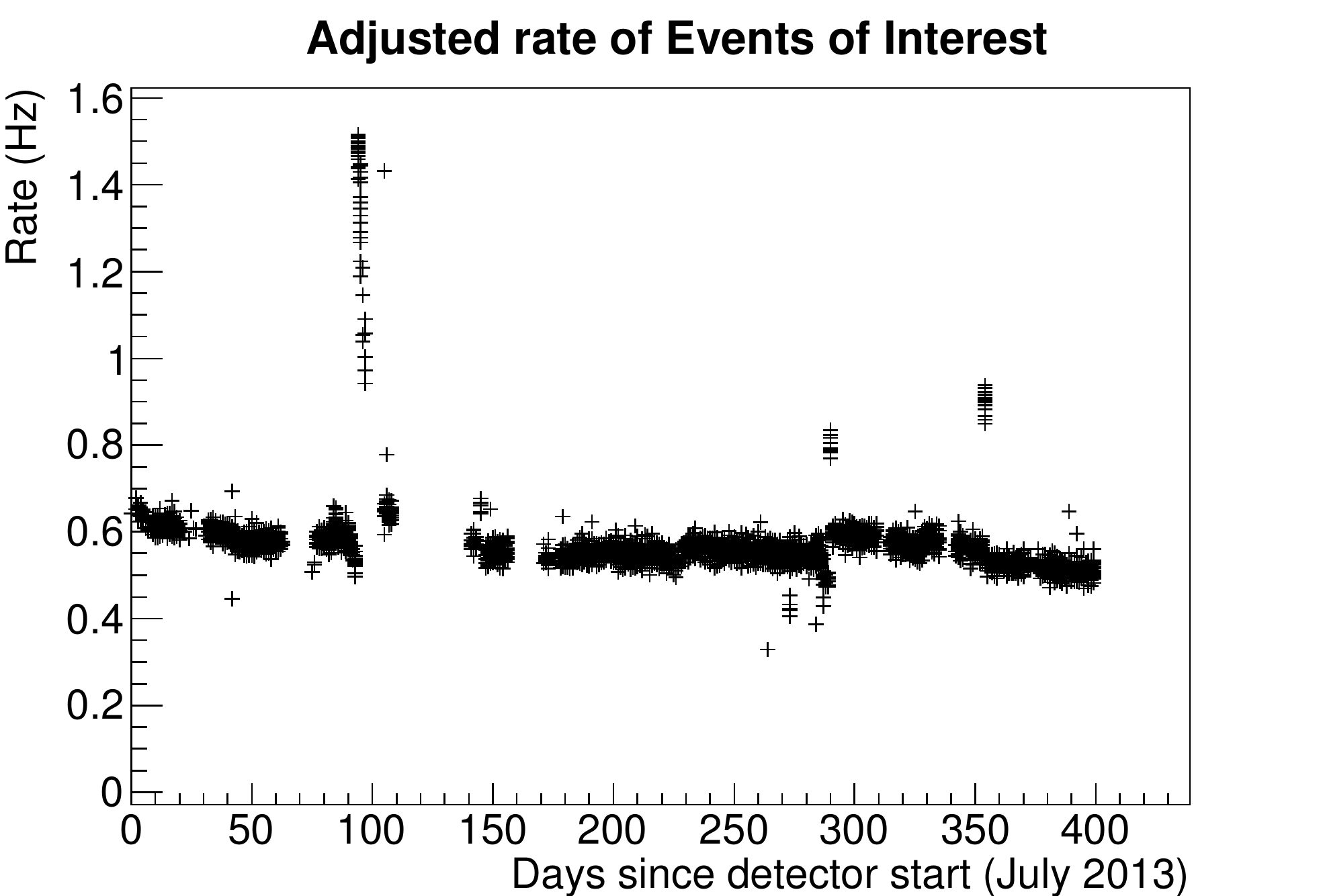}
\caption{The rate of ``Events of Interest'' following the application of adjusted energy cuts over the entire data taking period. Note that apart from some short term spikes, which were excluded from the analysis, the overall event rate is stable to within 12\%.}
\label{fig:eventRateAdjusted}
\end{figure}


\section{Calibrations, Signal Modeling and Analysis Criteria}
\label{sec:calibrations}
The WATCHBOY detector is calibrated with LEDs, permanently mounted throughout the target and veto regions, and a $^{252}$Cf fission source which can be positioned via a calibration tube next to the target. The LEDs produce single photoelectron pulses in the PMTs to allow matching of PMT gains and to determine the PMT anode charge per photoelectron. The $^{252}$Cf source allows us to determine the detector response to neutron captures on gadolinium. The response spectra are then used to tune a GEANT4-based~\cite{geant4} detector model, which is used to determine the optimal analysis cuts and corresponding efficiencies. Since the radionuclide sample of interest contains a final state neutron, this $8\,\mathrm{MeV}$ gamma ray cascade that results from neutron capture on gadolinium in the target is well-suited for the calibration.

The $^{252}$Cf source produces neutrons via spontaneous fission, along with a set of gamma-rays whose energies sum to an average of $6.65\,\mathrm{MeV}$~\cite{PhysRevC.87.024601}. The average neutron multiplicity per fission is $3.75$~\cite{Chadwick20112887,Reillyetal}. During a calibration run, if two or more correlated events are detected in the target, the first event may be due to gamma-rays arising from a fission, or a neutron if the gamma-rays were not detected. As described in Section 3, the second event in a correlated pair of events is highly likely to be a neutron capture, enabling the selection of a statistically pure spectrum of neutron capture events, which can be used to calibrate the detector response~\cite{Dazeley1,Dazeley2}. The $^{252}$Cf deployment tube is shown in yellow in Fig.~\ref{fig:watchboydesign}. The source is placed near the end of the calibration tube. The tube itself touches the target wall and defines the minimum distance of the source from the wall. During a deployment the distance from the source to the target wall is very stable at 3.5 cm.

As described above, the GEANT4-based Monte Carlo model was constructed to determine the predicted detector response for any event type, the best analysis cuts, and the resulting efficiencies. Three optical parameters - photon attenuation length, wall reflectivity, and PMT light collection efficiency - were tuned to produce a good match, in terms of spectral response and total light yield, between the model and the neutron capture spectrum derived from the $^{252}$Cf calibration.

The simulation tuning process was guided by initial boundary conditions. The average PMT quantum efficiencies were assumed to be slightly lower than the nominal factory specifications, since these are measured in ideal electromagnetic conditions and, for the purposes of simplicity, the model does not include the effect of various parts that support the PMT and mu-metal shielding, both of which obscure the view of each PMT to a small degree. Wall reflectivity is dominated by the sidewalls of the target, coated in highly reflective Teflon (GORE$\,\circledR$ DRP$\,\circledR$). The nominal reflectivity, in air, of this material is greater than $99\%$ in the blue and near UV. The top surface of the target was white polypropylene with an unknown, but certainly lower, reflectivity. The simulation model simply assumes constant overall wall reflectivity, which was varied during the tuning process over a range between $90\%$ and $99\%$. Finally, the water in WATCHBOY is not continually recycled or cleaned and it contains GdCl$_3$. Therefore, the photon attenuation length was assumed to be short relative to ideal water-based experiments such as SK, which has an attenuation length approaching $\sim100\,\mathrm{m}$~\cite{Fukuda:2002uc}.

Using neutron capture events from $^{252}$Cf calibration, we have two metrics to compare simulation and calibration. The first is the spectral response, which is used to adjust the simulated optical parameters to maximize agreement between the measured and simulated neutron capture energy spectra. The second is the proportion of light detected by each PMT on average in the target. Since the calibration source is positioned on the edge of the target, the mean fraction of light observed by each PMT is influenced by the PMT's position in the target and its proximity to the source.

Table~\ref{tab:TunedParameters} shows the central values and acceptable ranges for the attenuation length, reflectivity and light collection efficiency in the central detector, derived from a comparison of simulated and actual calibration data. We used the central values to fix our analysis cuts and predict the detection efficiency for the prompt beta and delayed neutron signatures of interest for our search. Using a $\chi^{2}$ test, we checked the sensitivity of the simulated spectra to parameter variations over a range of input parameters encompassing the central values. For small $\chi^{2}$ values, indicating good fits, the resulting efficiency variations formed a fairly tight range about the ideal (nominal) efficiency of approximately $\pm 1\%$. This uncertainty was included in the estimated overall efficiency uncertainties shown in Table~\ref{tab:analysiscuts}.

\begin{figure}[ht]
\centering 
\includegraphics[width=0.49\textwidth, height=0.25\textheight]{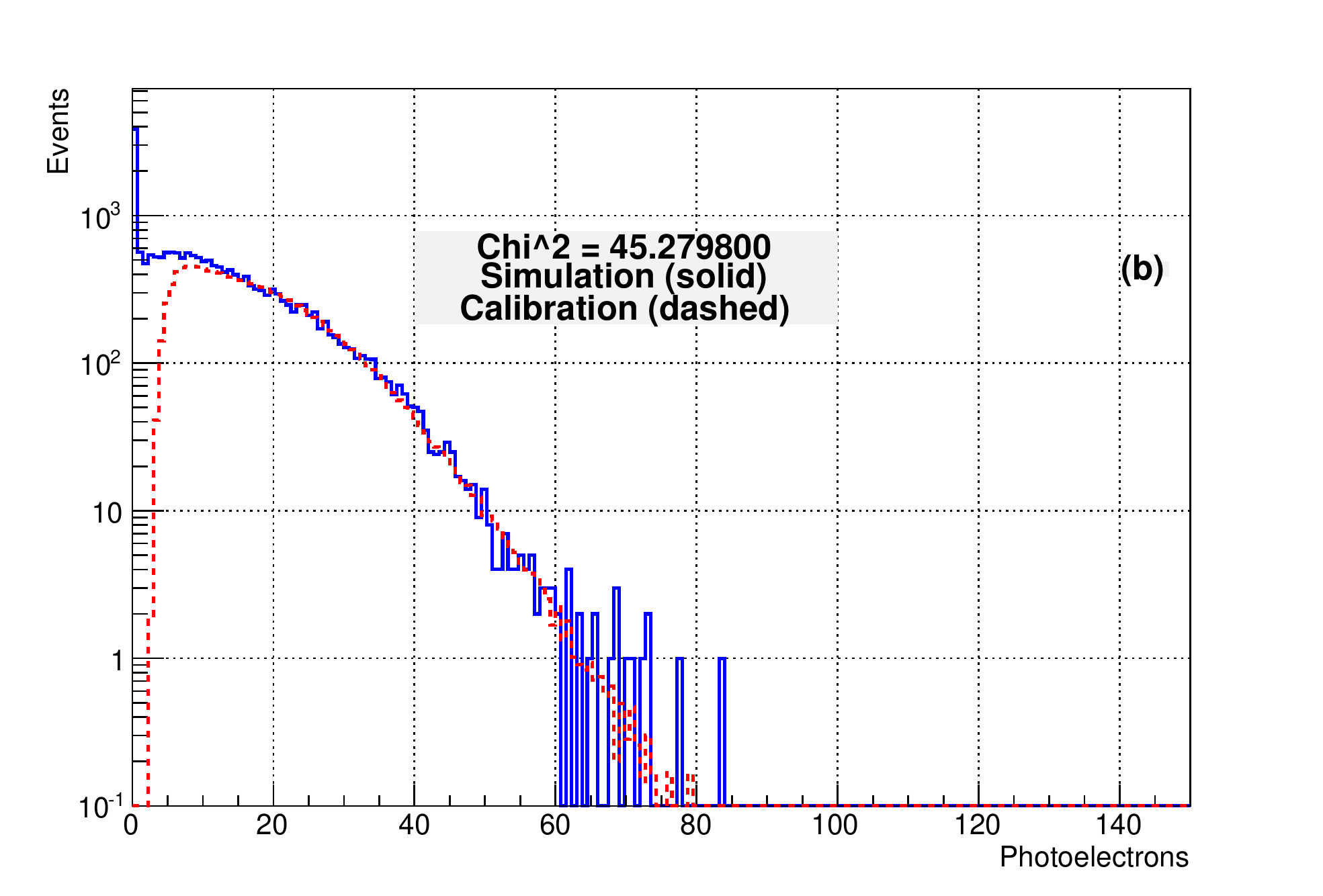}
\caption{\label{fig:2} 
A comparison of the detector spectral responses due to neutron captures in the target, obtained from a GEANT4 simulation after tuning (solid) and a $^{252}$Cf calibration run (dashed). Trigger thresholding, not included in the simulation, probably accounts for a large part of the difference between the curves below $\sim15\,\mathrm{pe}$.}
\end{figure}

\begin{table}[h]
\centering
\caption{The tuned photon transport parameters implemented in the GEANT4 detector model.}
\begin{tabular}{|c|c|}
\hline
Water attenuation length  & 24 meters \\
Wall reflectivity 			& 94\% \\
PMT Light collection efficiency & 20\% \\
 \hline
\end{tabular}
 \label{tab:TunedParameters}
 \end{table}

Fig.~\ref{fig:2} compares the neutron capture spectra from simulated and real (calibration) $^{252}$Cf data after tuning. Since the trigger is not included in the simulation, the two curves diverge below about $15\,\mathrm{pe}$. Above $15\,\mathrm{pe}$ the spectra match very well.

The tuned detector model was used to predict the response to $^{9}$Li decay throughout the target volume and then compared with the measured background spectrum. 
The $^{9}$Li signal can be separated into its $\beta$ and neutron components and compared 
with the background distribution to determine the optimal analysis cuts and evaluate the resulting efficiencies. The input energy spectrum~\cite{CaraThesis} 
for the $\beta$ decay has a Q value = 11.9 MeV. A mean neutron energy of $\sim 2\,\mathrm{MeV}$ was also used, though the neutron energies do not strongly 
influence the neutron capture detector response since the neutron usually thermalizes before capture. The top panels of Fig.~\ref{fig:analysiscuts} 
show the simulated spectra for the $\beta$s and the neutron captures that result from $^{9}$Li decays in the target.
\begin{figure*}[ht]
\centering
\includegraphics[width=0.49\textwidth, height=0.28\textheight]{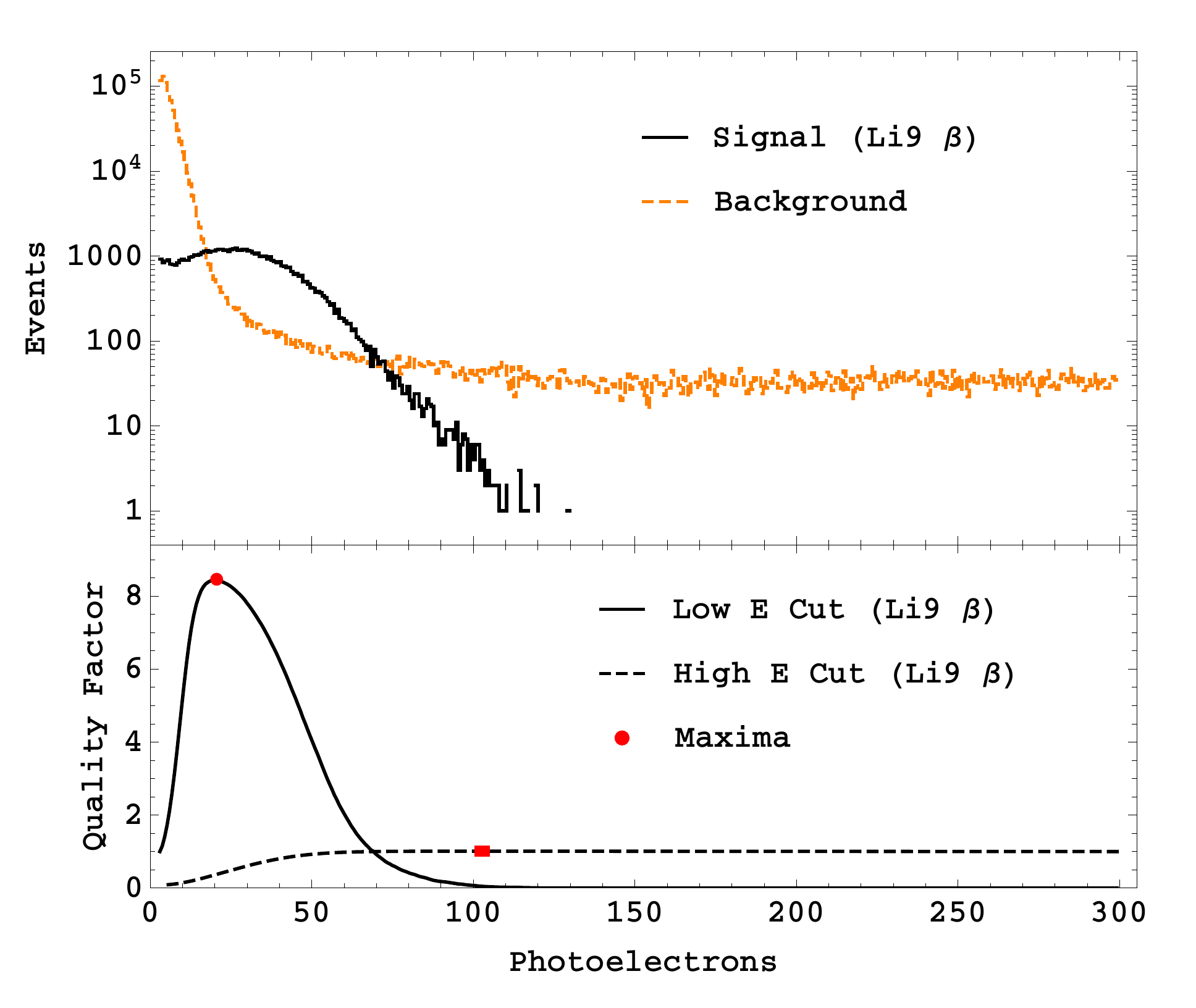}
\includegraphics[width=0.49\textwidth, height=0.28\textheight]{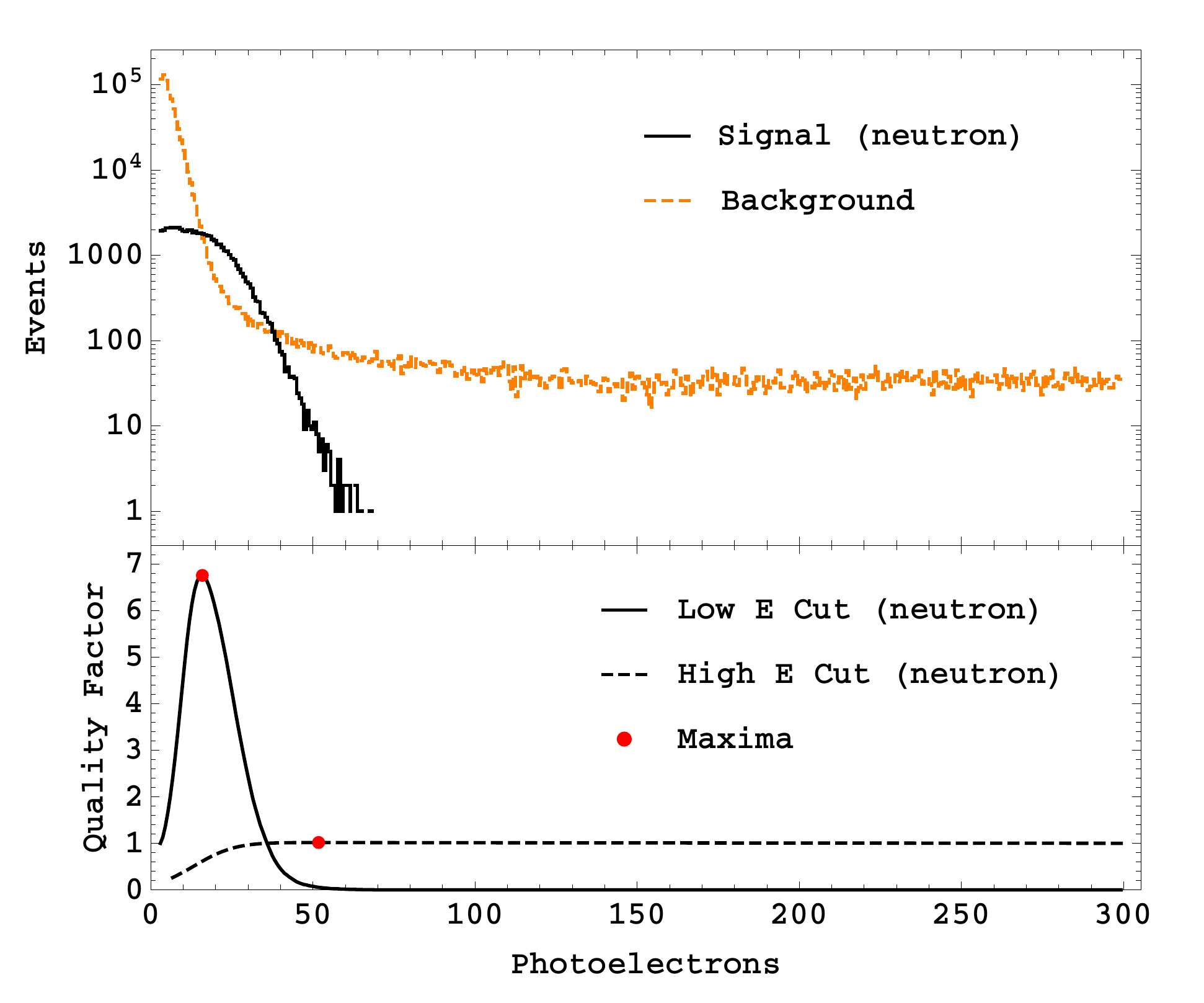}
\caption{\label{fig:analysiscuts} Top: the simulated detector response due to the $\beta$ (left) and the neutron (right) compared with the measured background. Bottom: the ``quality factor'' (defined in the text), that would result from an analysis cut at the corresponding number of detected photoelectrons. Higher quality factors following the imposition of a cut indicate improved signal efficiency and background rejection compared to cuts resulting in lower quality factors.}
\end{figure*}
\begin{table*}[ht]
\centering
\caption{\label{tab:analysiscuts} The analysis cuts and corresponding efficiencies determined from the detector simulation.  See text for further discussion of how the efficiencies were calculated.}
    \begin{tabular}{|c||c|c|c|}
        \hline
Source & Minimum & Maximum & Efficiency ($\%$) \\
\hline
$^{9}$Li $\beta$ & 20 pe & 100 pe & 62.8$\pm 9.4$ \\
\hline
Neutron & 16 pe & 53 pe & 42.2$\pm 6.3$ \\
\hline
$\beta$-Neutron inter-event time & 1 $\mu s$ & 200 ${\mu}$s & 95.8$\pm 3.6$\\
\hline
1ms muon veto analysis cut & & & 99.2 \\
\hline
Combined Total Efficiency & & & 25.2$\pm 5.4$\\
\hline
    \end{tabular}
\end{table*}
Also shown is the background spectral shape. Since the $\beta$ and neutron capture spectra are quite different from the background, we can maximize the statistical likelihood of detecting $^9$Li by an optimal choice of analysis cuts. All possible energy cuts, in terms of detected photoelectrons, were applied in turn, and the statistical advantage recalculated each time. The statistical advantage, represented by a ``quality factor'' at each possible cut, is calculated in a similar manner to Ref.~\cite{Sweany2011} and defined as:
\begin{equation}
Q_C = \bigg( \frac{S_C-B_C}{\sqrt{B_C}} \bigg) \bigg( \frac{\sqrt{B_T}}{S_T-B_T} \bigg)
\label{eq:Qfactor}
\end{equation}
where $S_C$ and $B_C$ correspond to the signal and background remaining after the cut respectively, while $S_T$ and $B_T$ are the corresponding values before the application of the cut. $Q_C$ is defined to be equal to $1$ if no cut is applied, or if there is no statistical advantage from the cut. {The best analysis cuts for the $\beta$ and the neutron, and the corresponding detection efficiencies, were evaluated by maximizing the quality factor. 

Timing cuts are also applicable since the $\beta$ and neutron are correlated. A $1-200\,\mu$s timing window was defined, with the lower limit set by the detector dead-time after a trigger, and the upper limit fixed large enough to ensure efficiency close to 100\%. 
The timing cuts were evaluated using the Monte Carlo and contributed an efficiency loss of $4.2 \pm 3.6\%$. There was a further efficiency loss of $0.8\%$ due to the 1 ms analysis veto which follows each muon in either the target or the veto. For the purposes of this analysis, a ``muon'' was defined as any event resulting in more than 100 photoelectrons in the target. A summary of all the $^{9}$Li analysis cuts is given in Table~\ref{tab:analysiscuts}. Note, the efficiencies for "$^{9}$Li $\beta$" and "Neutron" in Table~\ref{tab:analysiscuts} include the charge balance $< 0.6$ cut. From the simulation, approximately $6\%$ and $3.5\%$ of the $^{9}$Li $\beta$ and neutron capture events respectively are lost as a result of the charge balance cut, while the background is reduced by $95\%$. This cut therefore eliminates virtually all the events that originate from sources of electronic noise, while retaining nearly all the $^{9}$Li $\beta$-like or neutron-like candidates. In addition to the analysis cuts described above, if an ``Event of Interest'' occurs within 1 millisecond of a $^{9}$Li-like correlated pair, the multiplet of events is considered to be a correlated triple, disqualifying it as a $^{9}$Li candidate.

The uncertainty assigned to the neutron efficiency (in Table~\ref{tab:analysiscuts}) was estimated by comparing the simulation with $^{252}$Cf calibration data. Since the $^{252}$Cf source is not tagged, the calibration by itself does not provide an unambiguous measurement of the neutron efficiency. However, the inter-event time distribution obtained is highly sensitive to this efficiency. $^{252}$Cf fissions produce neutrons in bunches, with a known multiplicity distribution of mean 3.75 ~\cite{Reillyetal}. The time intervals between the neutrons detected in each bunch form an inter-event time distribution with a correlated and an uncorrelated component. The magnitude of the correlated component, the exponential slope of the uncorrelated component, and the observed multiplicity distribution are particularly sensitive to neutron efficiency.  A simple neutron capture timing model, consisting of the sum of three independent exponential terms, was used to model the inter-event time distribution. The three terms represent the following effects:
\begin{enumerate}
	\item  Thermal neutron capture in the target
	\item  The thermal capture time behavior of neutrons caught in the veto, but transition into the target
	\item  A subtracted component representing the thermalization of $\sim$ 2 MeV neutrons in water
\end{enumerate}
The terms above describe the probability of capture of each neutron following emission from the source as a function of time in this model. Detection depends upon the efficiency. Prompt fission gamma-rays, if detected, occur at t=0 by definition. Fission delayed gamma-rays can occur at any time. The fission rate of the source in the model was fixed at 564 Hz, since it was measured independently at NIST to an accuracy of 1.6\% \cite{NIST}. 
The parameter space defined by a range of neutron efficiencies, prompt fission gamma-ray efficiencies, and delayed gamma-ray efficiencies was scanned to identify values that reproduced both the observed inter-event time and neutron multiplicity distributions. A number of viable solutions were found - all with neutron efficiencies in the range $10.5\%$ to $11.5\%$. The best solution overall was found at $11.0\%$. The predicted neutron efficiency from the simulation was $12.6\%$, almost $15\%$ higher. We therefore assigned an uncertainty of $\pm 15\%$. We note that the observed 15\% difference can be accounted for by including a small uncertainty in the source position relative to the target wall.

The uncertainty in the neutron detection efficiency arises from three independent processes - the fraction of neutrons from the calibration source that capture in the target, the fraction of neutron capture gamma rays that Compton scatter inside the target, and the uncertainty in photon production from scattered $\beta$ particles. If we assume, conservatively, that the observed uncertainty was entirely due to uncertainties in photon production, rather than neutron transport or Compton scattering, then the $^{9}$Li $\beta$ uncertainty can also be assigned the same value. We stress that this is a very conservative assumption, enforced by the lack of suitable $\sim10$ MeV calibration sources. If the neutron transport or Compton scattering uncertainties were to contribute significantly to the total, the photon production component, which forms the major part of the $^{9}$Li $\beta$ uncertainty, would be reduced. New calibration sources will be deployed in the near future in an attempt to reduce this uncertainty for future publications. For now we assigned the conservative value ($\pm 15\%$) for $^{9}$Li $\beta$ detection also.


While the primary objective of WATCHBOY was to measure $\beta$-neutron emitting radionuclides, the detection of single $\beta$ emitting radionuclides provides an important additional check on the radionuclide detection capabilities of the detector. Our data show robust evidence for these single $\beta$ events. The ability to identify them helps confirm WATCHBOY's suitability for isolating the $^{9}$Li $\beta$-neutron signature. In \ref{app:appendix} we present a method for tagging so-called ``showering'' muons and study the resulting single-$\beta$ radionuclide production for WATCHBOY. A more complete analysis, including efficiency estimates, will be presented in a separate paper.

\section{$^9$Li $\beta$-Neutron Result}
\begin{figure}[ht]
\centering
\includegraphics[width=0.50\textwidth, height=0.25\textheight]{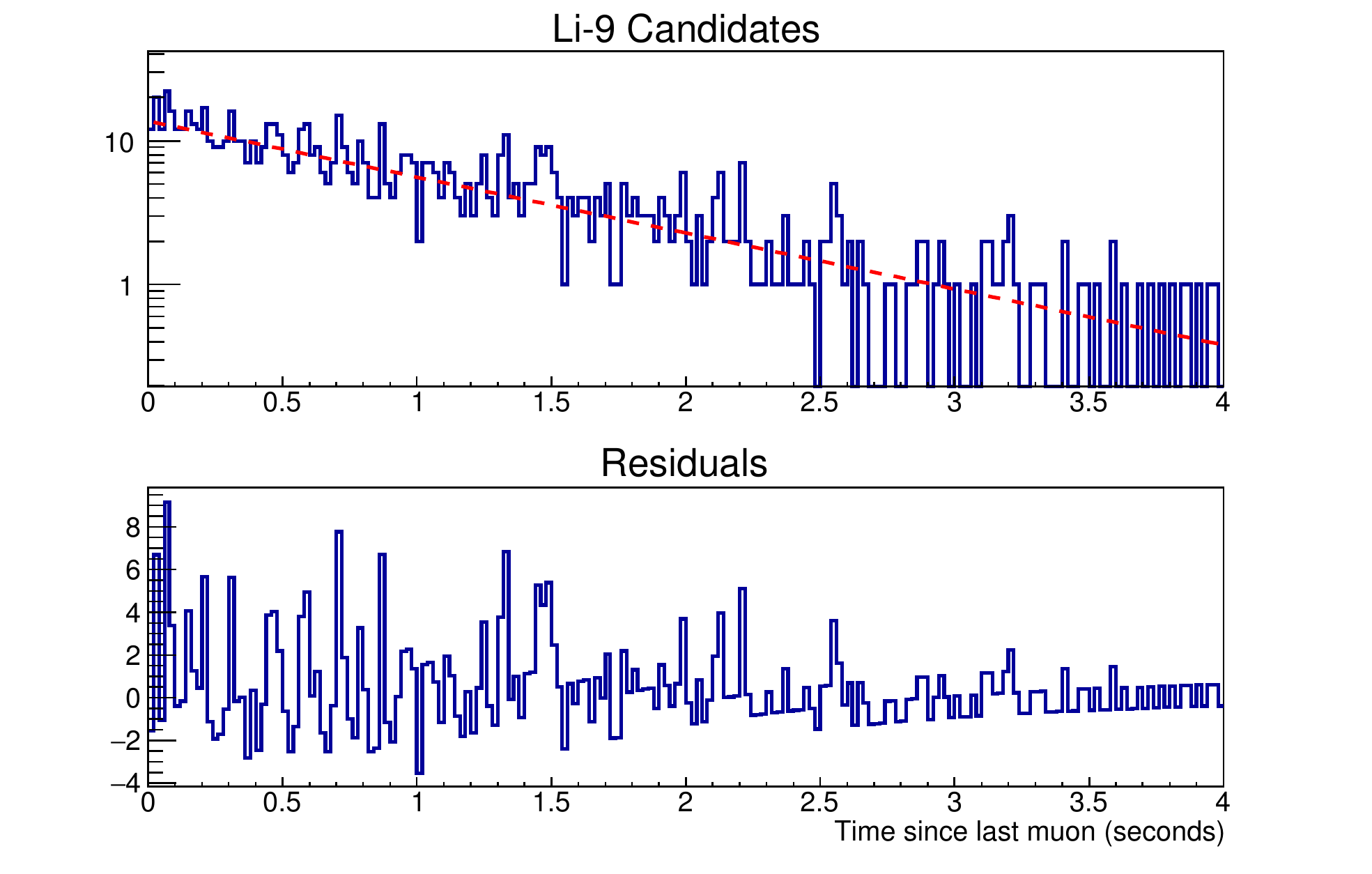}
\caption{\label{fig:residuals} A plot of the elapsed time since last muon through the target for the entire set of $^9$Li candidates (top). The $^9$Li candidates are correlated event pairs that satisfy the charge requirements of Table~\ref{tab:analysiscuts}. Below are the residuals after subtracting the exponential, shown as a dashed line.}
\end{figure}
\begin{figure}[ht]
\centering
\includegraphics[width=0.49\textwidth, height=0.25\textheight]{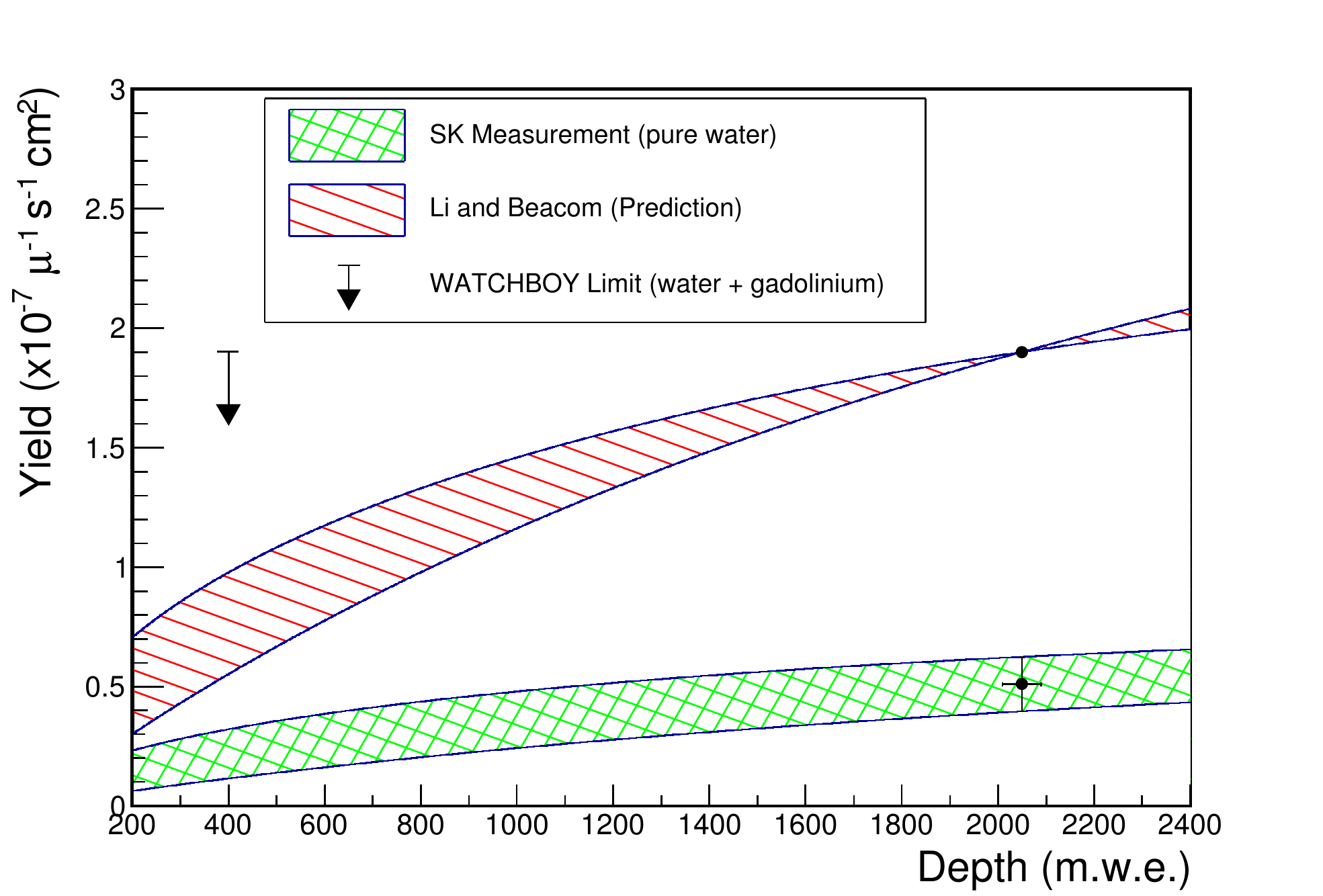}
\caption{\label{fig:depthScaling} The 90\% upper limit on $^{9}$Li production from this work compared to the recent SK measurement~\cite{SKRadNucs2015} and the prediction of Li and Beacom~\cite{Li:2015kpa}. The measurements and predictions, both at the SK depth, were scaled with depth according to the mean muon energy, shown as the hashed and striped regions. The allowed range of scaling factors were from Hagner et al.~\cite{Hagner2000} ($<E_{\mu}>^{0.5}$ to $<E_{\mu}>^{0.93}$).}
\end{figure}

We now describe the $^9$Li $\beta$-neutron analysis. As described in Section 2, a $^9$Li decay candidate in WATCHBOY is defined as a pair of correlated events, consistent with a $\beta$-neutron pair, correlated with an earlier showering or non showering muon passing through the target. The $\beta$-neutron candidates must have higher energy than typical background, with charge balance values consistent with relatively even light distribution among the PMTs. A target muon is any target event producing more signal than the upper energy limit of $^{9}$Li ($> 100$ photoelectrons). The first of the correlated events must have total charge in the target consistent with the $^9$Li $\beta$ of Table~\ref{tab:analysiscuts}, the second consistent with neutron capture on gadolinium. We further require that the correlated pair not occur within $1\,\mathrm{ms}$ of any other ``event of interest'' in the target, or a muon in either the target or veto, which excludes spallation neutrons. In Fig.~\ref{fig:residuals} (top), the time to the most recent target muon is plotted for all of the remaining $^9$Li $\beta$-neutron candidates. If $^9$Li forms a significant proportion of these events, we expect to see a subset of events correlated with target muons with a lifetime consistent with the lifetime of $^9$Li ($257\,\mathrm{ms}$). In Fig.~\ref{fig:residuals} (bottom), the residuals are plotted. The magnitude of the observed $^9$Li component ($20.2\pm25.4$ events) was calculated from a double exponential fit to Fig.~\ref{fig:residuals} (top), fixing the slopes to the known $^9$Li and background exponentials {(the target muon rate is 0.9 Hz)}, while allowing the magnitudes to float. No significant excess consistent with $^{9}$Li decay is observed in this data set. The observed statistical significance of the $^9$Li component was therefore $+0.8\sigma$. The total number of background event pairs recorded over 207 live days was 799. Given these results the $90\%$ upper limit on the total number of $^{9}$Li created in the target was 266 assuming a detection efficiency for the $\beta$-neutron branching ratio of $25.2\pm5.4\%$. From these 
results we can calculate the upper limit on the $^{9}$Li yield using the relation
\begin{equation}
Y < \frac{N}{\epsilon B_{\beta n} T L_{\mu} R_{\mu} \rho}  = \frac{266}{T L_{\mu} R_{\mu} \rho}
\label{eq:YieldCalc}
\end{equation}
where $N$ is the upper limit on the number of observed $^{9}$Li candidates, $\epsilon$ is the efficiency, $B_{\beta n}$ is the $^{9}$Li $\beta$-neutron branching ratio ($50.8 \pm 0.9 \%$), $T$ is the livetime ($17.9 \times 10^6$ seconds), $L_{\mu}$ the average muon path length through the target ($87.0 \pm 3.1$cm), $R_{\mu}$ the muon rate ($0.9 \pm 0.007$ Hz) and $\rho$ the target density(1.0 g/cm$^3$). The average muon path length was calculated from the muon angular distribution at WATCHBOY, determined using the muon flux parametrization as a function of depth and $\theta$ defined by Mengyun et al.~\cite{mengyun}. The uncertainty on the rate was determined using the GEANT4 simulation assuming the 15\% uncertainty on Cherenkov photon production and the $\cos \theta$ distribution calculated earlier. The 90\% upper limit on the $\beta$-neutron branching ratio of $^{9}$Li at the WATCHBOY depth (400 m.w.e.) is therefore $0.97 \times 10^{-7} \mu^{-1} \mathrm{g}^{-1} \mathrm{cm}^{2}$. This translates to an upper limit on total $^{9}$Li production of $1.9 \times 10^{-7} \mu^{-1} \mathrm{g}^{-1} \mathrm{cm}^{2}$.

The WATCHBOY 90\% upper limit on the yield is shown in Fig.~\ref{fig:depthScaling} together with the $^{9}$Li prediction of Li and Beacom~\cite{Li:2014}, the recent measurement from SK~\cite{SKRadNucs2015} and their expected yield as a function of depth. Both the Li and Beacom and SK depth projection include a scaling uncertainty with respect to average muon energy (depth) using the range of acceptable scaling factors ($<E_{\mu}>^{0.5}$ to $<E_{\mu}>^{0.93}$) given by Hagner et al.~\cite{Hagner2000}. The SK measurement is seemingly in conflict with the Li and Beacom prediction by a factor of $\sim$4, though factor $\sim$2 disagreements are common in liquid scintillator ~\cite{KamLANDbackgrounds}. WATCHBOY, while not yet able to distinguish between these results due to the small data set, is a more efficient detector due to the capability to tag neutron captures on gadolinium.  This proof of concept suggests that larger Gd-doped water detectors should be able to efficiently and definitively measure $\beta$-neutron radionuclide rates, and, if deployed at a different depth than SK, help reduce uncertainties in the yield scaling factors between different depths.}

Extrapolating the WATCHBOY result to the WATCHMAN depth (1500 m.w.e.), we obtain an upper limit on the $^{9}$Li yield of between $3.3 - 5.4 \times 10^{-7} \mu^{-1} \mathrm{g}^{-1} \mathrm{cm}^{2}$. This corresponds to a rate of  between 13 and 21 detected $^{9}$Li per kiloton per day assuming 50\% detection efficiency and accounting for the $\beta$-neutron branching ratio. The KamLAND and SK ~\cite{Eguchi:2002dm,KamLANDOscillation,KamLANDbackgrounds,Li:2015kpa} detectors have both achieved $> \sim 90\%$ $^{9}$Li veto efficiency (KamLAND $99\%$) by vetoing events within 2 seconds and 3 meters of each muon track. At $90\%$ efficiency, the $^{9}$Li background contribution to the antineutrino signal at WATCHMAN would be less than $\sim 2$ events per day, sufficiently low to allow for the successful detection of antineutrinos from the nearby Perry reactor.

\section{Conclusion}
We have presented the first results of the WATCHBOY detector, {designed to determine} the rate of cosmogenically produced radionuclides, such {as the $\beta$-neutron emitting isotopes $^9$Li and $^{8}$He, that may} act as backgrounds {in} future water-based antineutrino detectors. The {$^{9}$Li detection efficiency was determined via a GEANT4 simulation, which was tuned from calibrations.} {From 207 live days, the WATCHBOY detector observed $20.2\pm25.4$ $^{9}$Li events. The 90\% upper limit on the yield of $^{9}$Li , derived from these results was $1.9 \times 10^{-7} \mu^{-1} \mathrm{g}^{-1} \mathrm{cm}^{2}$, at a depth of approximately 400 m.w.e. The primary source of backgrounds in WATCHBOY were correlated pairs of neutron captures, initiated by fast neutrons from the surrounding rock. WATCHBOY observes a total of 799 of these events, at a rate of 3.9 per day. In addition, WATCHBOY observes a clear excess of single $\beta$ emitting radionuclide candidates following showering muons, where showering muons are defined as muons followed by at least two correlated neutron captures (see \ref{app:appendix}). Though these single $\beta$ emitting radionuclides are not antineutrino detector backgrounds, their detection does serve as a verification of the detector's capabilities. A more complete study, including a determination of the total radionuclide yield following showering muons, will be provided in a later paper. In the near future the rate of correlated neutron backgrounds will be compared with measurements of the fast neutron energy spectrum currently being completed by the WATCHBOY\textquotesingle s sister project called MARS~\cite{MARSDetector} (Multiplicity and Recoil Spectrometer), as a cross check. The predicted rate of $^{9}$Li backgrounds at WATCHMAN, based on these measurements is less than $\sim 2$ events per day, sufficiently low to allow for the successful detection of antineutrinos from the nearby Perry reactor.

\section*{Acknowledgments}
The authors would like to thank John Steele for programming the trigger logic in the FPGA, Doug Dobie and Kevin Morris for the engineering design, John Bower and Darrell Carter for valuable help with detector construction.
This work was performed under the auspices of the US
Department of Energy by Lawrence Livermore National Laboratory under Contract DE-AC52-07NA27344, release no. LLNL-JRNL-677613. This work was supported by the U.S. Department of Energy Office of Defense Nuclear Nonproliferation Research and Development.

\appendix
\section{Single $\beta$ Production Following Muon-Induced Hadronic Showers}
\label{app:appendix}
Li and Beacom~\cite{Li:2015kpa} assert that essentially all radionuclides in the SK (undoped) water Cherenkov detector are produced during muon-induced hadronic showers. A ``showering'' muon is characterized by multiple hadronic interactions, which generate multiple neutrons and pions along the muon's track. In addition, certain radionuclide production processes themselves create secondary neutrons. Based on these assumptions, we identify showering muons in WATCHBOY using the following method: if two or more neutron-like events occur in the target within $1\,\mathrm{ms}$ of a muon, the muon is tagged as a showering muon and subsequent $\beta$ events (at times greater than $1\,\mathrm{ms}$) are identified. The muon may traverse either the target or the veto. We require at least two neutron-like events to reduce the likelihood of accidental coincidences and ensure a clean sample of showering muons. The reduction in background that results from this requirement is illustrated in Fig.~\ref{fig:muonshowertagging}. If the claim is true, an excess of single $\beta$-like events correlated with showering muons should be observed.
\begin{figure}[ht]
\centering
\includegraphics[width=0.49\textwidth, height=0.25\textheight]{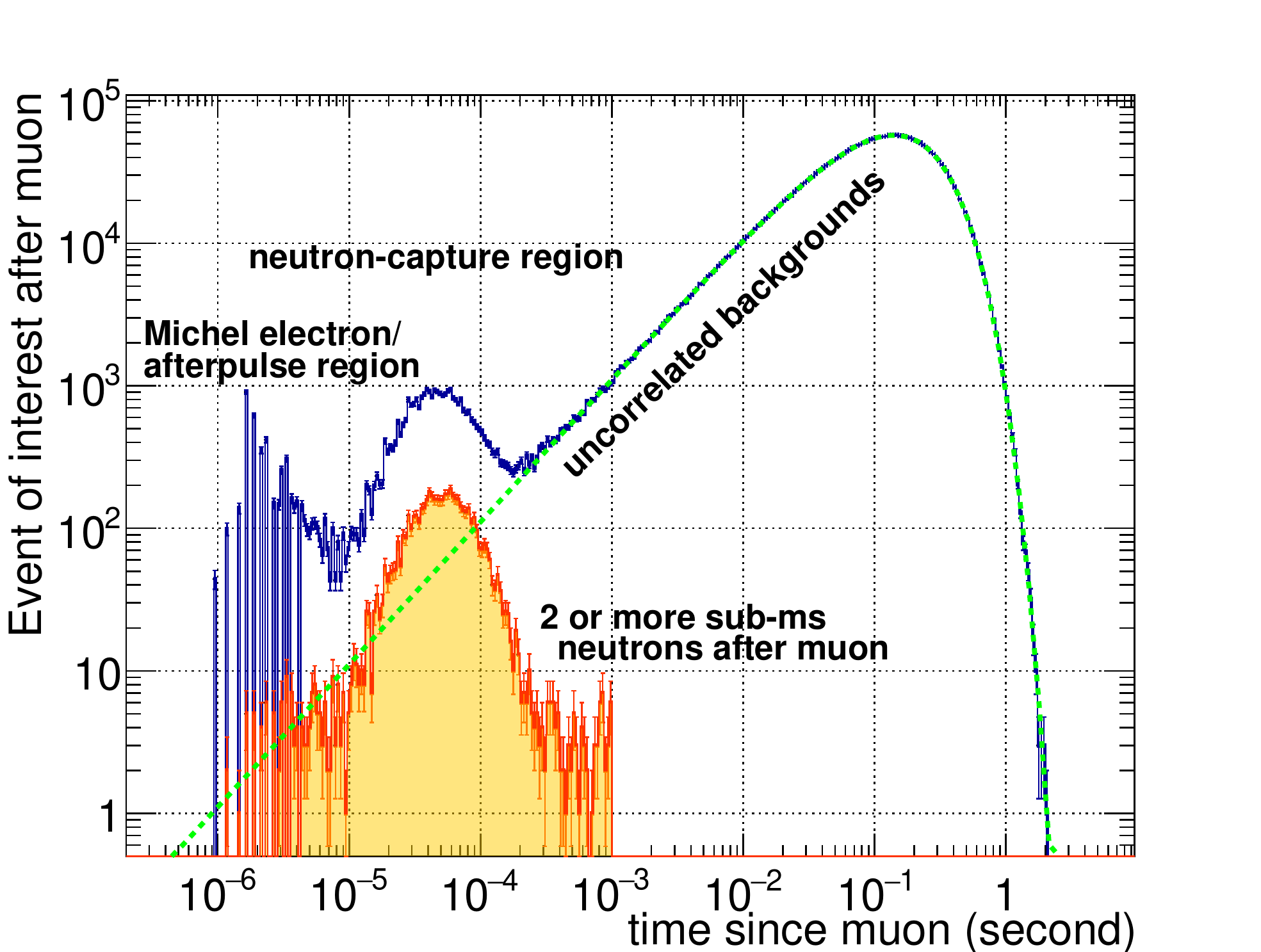}
\caption{\label{fig:muonshowertagging} {The distribution of event times following muons in WATCHBOY. A showering muon is identified by the presence of 2 or more neutron-like events within $1\,\mathrm{ms}$. This tagging procedure allows the removal of nearly all backgrounds due to pile up of other muons. The uncorrelated events were fitted in the region of 1 ms to 2 second, {where good agreement is observed between the data and the uncorrelated expectation.}}}
\end{figure}
We observe an average of $9.8$ showering muons per day tagged in this fashion. Because of the low rate of these muons, a good separation between correlated events and non-correlated events can be achieved, as is shown in Fig.~\ref{fig:muonshowertagging} and Fig.~\ref{fig:muonshowerexcess}.  Tagging random non-showering muons was done as a crosscheck on the potentially correlated $\beta$ spectrum; this spectrum is shown in Fig.~\ref{fig:muonshowerexcess}. There is a selection of events that show correlation to showering muons, revealed as the excess when comparing the uncorrelated and correlated spectra. In addition to the correlated events shown, we include the time profile of a GEANT4 simulation of neutron capture events throughout the target assuming a flat input neutron energy spectrum extending up to 1 GeV. These events follow an exponential curve consistent with neutron thermalization and capture on gadolinium. The measured and simulated time profiles are consistent above $\sim 30 \mu s$. Below $30 \mu s$ the number of detected correlated events is below expectation. The difference is due to the presence of baseline variability which tends to follow extremely bright events, such as muons. Baseline variability is revealed by anomalously high or low baseline levels immediately preceding a physics event (a sample of the baseline charge of every PMT is taken for approximately $100 ns$ prior to every trigger). Events which show evidence of baseline variability are identified and eliminated, leading to some loss of efficiency following muons.
\begin{figure}[ht]
\centering
\includegraphics[width=0.49\textwidth, height=0.25\textheight]{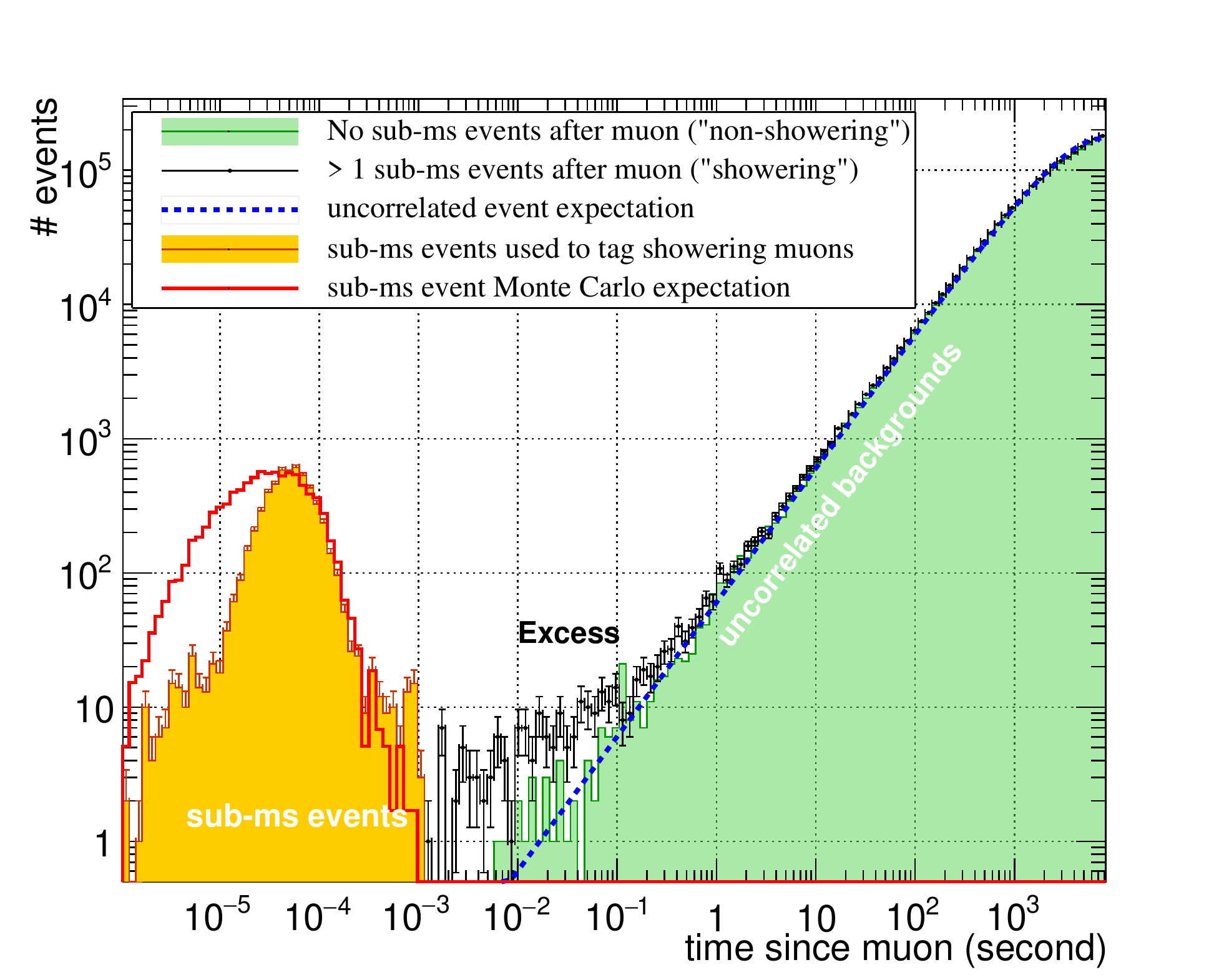}
\caption{\label{fig:muonshowerexcess} Time of events of interest with respect to both showering and randomly selected muons. Radionuclides are expected in the $1\,\mathrm{ms}$ to $10\,\mathrm{s}$ range. We observe an excess of events in a time window between $1\,\mathrm{ms}$ and $1\,\mathrm{s}$. The sub-ms time scale muogenic particles defining the presence of a shower are shown as the shaded distribution to the far left. The expected time signature of sub-ms muogenic particles was evaluated with Monte Carlo and is shown as a solid line, the measurement is attributed to a delay in baseline recovery following a muon.}
\end{figure}

In addition to the time-correlations in the showering muon population, Fig.~\ref{fig:highenergyexcess} shows that the energy spectra of the showering and non-showering samples differ significantly. A surplus of events at higher energy is observed in the showering muon sample, consistent with a signature from $\beta$ emission of radionuclides. Fig.~\ref{fig:highenergyexcess} (bottom) also shows the energy distribution of the excess events subtracted by {a normalized selection of} uncorrelated events found in the $1\,\mathrm{ms}$ to $1\,\mathrm{second}$ time window.

Since this paper focuses on the time-correlated $\beta$-neutron background to reactor anti-neutrino detection, further analysis of the single $\beta$ emitter is relegated to a separate paper. Summarizing this analysis, we have observed an excess of events consistent with single-$\beta$ radionuclide production, correlated to muon induced showers. This result supports the hypothesis advanced by Li and Beacom~\cite{Li:2015kpa}, and helps confirm the ability of WATCHBOY to trigger on the closely analogous $\beta$-neutron radionuclide population.
\begin{figure}[ht]
\centering
\includegraphics[width=0.49\textwidth, height=0.25\textheight]{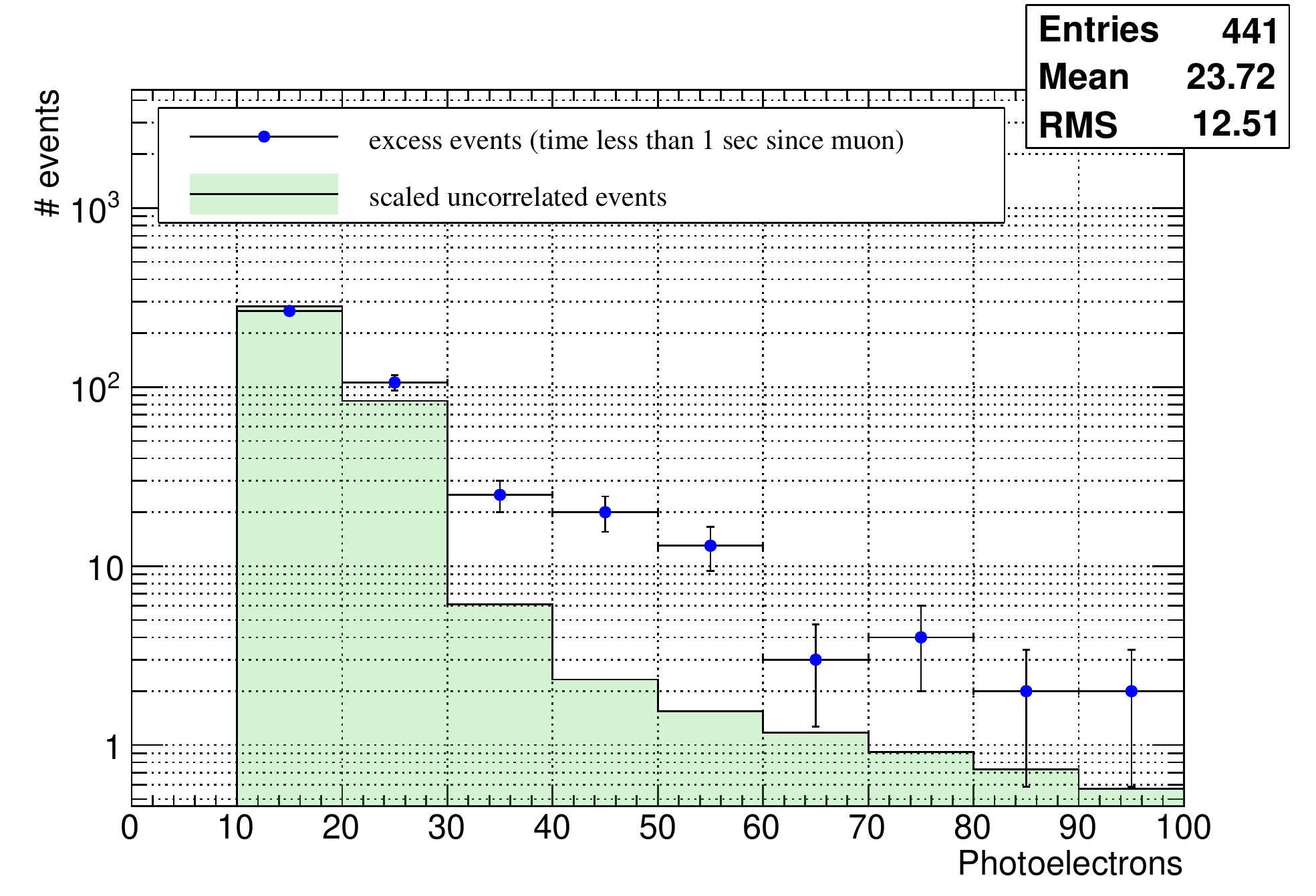}
\includegraphics[width=0.49\textwidth, height=0.25\textheight]{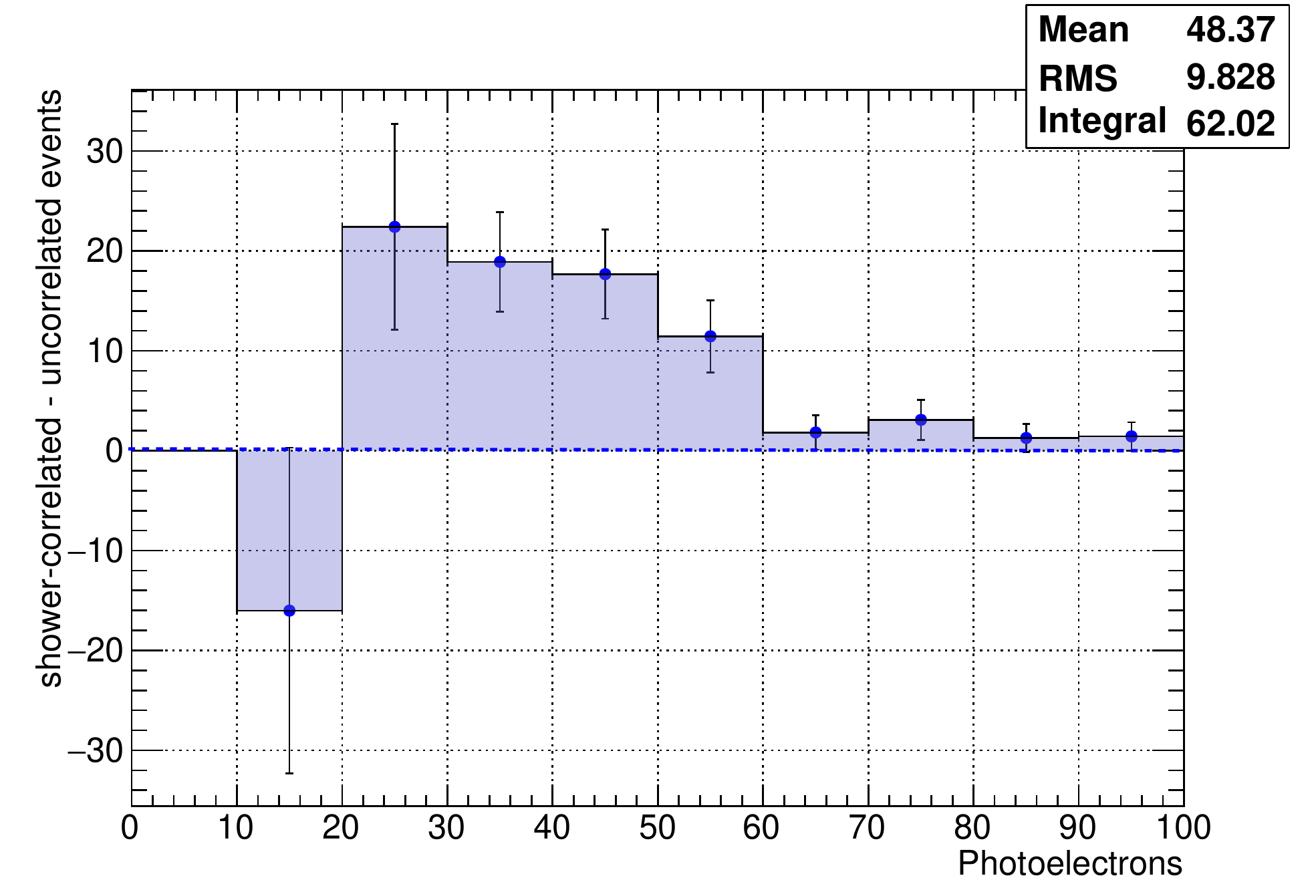}
\caption{\label{fig:highenergyexcess} The top panel shows the energy spectra of correlated and non-correlated events within $1\,\mathrm{second}$ of a {showering} muon compared to the uncorrelated/accidental background. The uncorrelated spectrum was normalized by the number of uncorrelated events expected from the fit shown on Fig.~\ref{fig:muonshowerexcess}. The bottom panel shows the background subtracted energy distribution of the excess events. {This forms the single-$\beta$ radionuclide candidate spectrum in WATCHBOY.}}
\end{figure}


\clearpage
\bibliography{WATCHBOYAnalysis}

\end{document}